\documentclass[twocolumn,showpacs,preprintnumbers,amsmatsh,amssymb,aps,prc]
{revtex4-1}
\usepackage{graphicx}
\graphicspath{{.}}
\usepackage{dcolumn}
\usepackage{bm}
\usepackage{float}
\usepackage[Gray,squaren,thinqspace,thinspace]{SIunits} 
\usepackage{amsmath}

\usepackage{grffile}
\newcommand{\ket}[1]{\mid#1\rangle}
\newcommand{\bra}[1]{\langle#1\mid}

\newcommand{\braket}[2]{\langle#1\mid#2\rangle}
\begin{document}

\title{Factorization of exclusive electroinduced two-nucleon knockout}
\author{Camille Colle}
\email{Camille.Colle@UGent.be}
\author{Wim Cosyn}
\email{Wim.Cosyn@UGent.be}
\author{Jan Ryckebusch}
\email{Jan.Ryckebusch@UGent.be}
\author{Maarten Vanhalst}
\email{Maarten.Vanhalst@UGent.be}
\affiliation{Department of Physics and Astronomy,\\
 Ghent University, Proeftuinstraat 86, B-9000 Gent, Belgium}
\date{\today}
\begin{abstract}
We investigate the factorization properties of the exclusive
electroinduced two-nucleon knockout reaction $A(e,e'pN)$.  A
factorized expression for the cross section is derived and the
conditions for factorization are studied. The $A(e,e'pN)$ cross
section is shown to be proportional to the conditional center-of-mass
(c.m.)  momentum distribution for close-proximity pairs in a state
with zero relative orbital momentum and zero radial quantum number.
The width of this conditional c.m.~momentum distribution is larger
than the one corresponding with the full c.m.~momentum distribution.
It is shown that the final-state interactions (FSIs) only moderately
affect the shape of the factorization function for the $A(e,e'pN)$
cross sections. Another prediction of the proposed factorization is
that the mass dependence of the $A(e,e'pp)$ $[A(e,e'pn)]$ cross
sections is much softer than $\frac{Z(Z-1)}{2}$ $[NZ]$.
\end{abstract}
%

\maketitle
\section{Introduction} 
\label{sec:intro}

In recent years, substantial progress has been made in exploring the
dynamics of short-range correlations (SRCs) in nuclei.  On the
experimental side, exclusive $A(p,2p+n)$ \cite{Tang:2002ww} and
$A(e,e'pN)$ \cite{Niyazov:2003zr,Shneor:2007tu,Subedi:2008zz}
measurements have probed correlated pairs in nuclei and identified
proton-neutron (pn) pairs as the dominant contribution.  Inclusive
$A(e,e')$ \cite{Egiyan:2003vg,Egiyan:2005hs,Fomin:2011ng} measurements
in kinematics favoring correlated pair knockout, have provided access
to the mass dependence of the amount of correlated pairs relative to
the deuteron.  On the theoretical side, \textit{ab initio}
\cite{Schiavilla:2006xx,Wiringa:2008vv,Feldmeier:2011tt,Wiringa:2013ala},
cluster expansion \cite{Alvioli:2007zz,Alvioli:2012dd,Alvioli:2013ff},
correlated basis function theory
\cite{AriasdeSaavedra:2007qg,Bisconti:2007dd}, and low-momentum
effective theory \cite{Bogner:2012zm}, calculations have provided
insight in the fat high-momentum tails of the momentum distributions
attributable to multinucleon correlations. Tensor correlations have
been identified as the driving mechanism for the fat tails just above
the Fermi momentum. The highest momenta in the tail of the momentum
distribution are associated with the short-distance repulsive part of
the nucleon-nucleon force and $N \ge 3$ correlations.  Recent reviews
of nuclear SRC can be found in
Refs.~\cite{Arrington:2011xs,Frankfurt:2008zv}.

We have proposed a method to quantify the amount of correlated pairs
in an arbitrary nucleus
\cite{Vanhalst:2011es,Vanhalst:2012ur,Vanhalst:2012zt}.  Thereby, we
start from a picture of a correlated nuclear wave function as a
product of a correlation operator acting on an independent-particle
model (IPM) Slater determinant $\Psi _{A}^\text{IPM}$ \cite{Bogner:2012zm}.
The SRC-susceptible pairs are identified by selecting those parts of
$\Psi _{A}^\text{IPM}$ that provide the largest contribution when subjected
to typical nuclear correlation operators.  It is found that IPM
nucleon-nucleon pairs with vanishing relative orbital momentum and
vanishing relative radial quantum numbers, receive the largest
corrections from the correlation operators. This can be readily
understood by realizing that IPM close-proximity pairs are highly
susceptible to SRC corrections. This imposes constraints on the
relative orbital and radial quantum numbers of the two-nucleon cluster
components in the IPM wave functions which receive SRC corrections.

With the proposed method of quantifying SRC we can reasonably account
for the mass dependence of the $\frac{ A(e,e') }{d(e,e')}$ ratio under
conditions of suppressed one-body contributions (Bjorken $x_B \gtrsim
1.2$) \cite{Vanhalst:2012ur} and the mass dependence of the magnitude
of the EMC effect \cite{Vanhalst:2012zt,WIMINPC2013}. In connecting the SRC
information to inclusive electron-scattering data at Bjorken $x_B 
\gtrsim 1.2$, there are complicating issues like the role of c.m. motion
\cite{Arrington:2012ax,Vanhalst:2012ur} and final-state interactions
(FSIs) \cite{Benhar:2013dq}. More quantitative information on SRC and
their mass and isospin dependence, is expected to come from exclusive
electroinduced two-nucleon knockout which is the real fingerprint of
nuclear SRC \cite{Starink2000}.  Reactions of this type are
under investigation at Jefferson Laboratory (JLab) and results for
$^{12}$C$(e,e^{\prime}pN)$ have been
published~\cite{Shneor:2007tu,Subedi:2008zz}.

In this paper, we investigate the factorization properties of the
exclusive $A(e,e'pN)$ reaction. Factorization is a particular result
that emerges only under specific assumptions in the description of the
scattering process. It results in an approximate expression for the
cross section which becomes proportional to a specific function of selected
dynamic variables. For exclusive quasielastic $A(e,e'p)$ processes,
for example, the factorization function is the one-nucleon momentum
distribution evaluated at the initial nucleon's momentum. It will be
shown that for exclusive $A(e,e'pN)$ these roles are respectively
played by the c.m.~momentum distribution for close-proximity pairs and
the c.m.~momentum of the initial pair.

In Sec.~\ref{sec:momdistr} we present calculations for the pair
c.m. momentum distribution in the IPM. It is shown that the
correlation-susceptible IPM pairs have a broader c.m. width than those
that are less prone to SRC corrections.  In Sec.~\ref{sec:fac}, we
show that after making a number of reasonable assumptions, the
eightfold $A(e,e'pN)$ cross section factorizes with the conditional
pair c.m. momentum distribution as the factorization function.  In
Sec.~\ref{sec:mc} we report on results of Monte Carlo simulations for
$A(e,e'pp)$ processes in kinematics corresponding to those accessible
in the JLab Hall A and Hall B detectors. We study the effect of
typically applied cuts on several quantities.  In Sec.~\ref{sec:FSI}
it is investigated to what extent FSIs affect the factorization function of
the exclusive $A(e,e'pN)$ process. Finally, our conclusions are stated
in Sec.~\ref{sec:concl}.


\section{Pair Center-of-mass momentum distributions} \label{sec:momdistr}

In this section we study the pp and pn pair c.m. momentum distribution for
$^{12}$C, $^{27}$Al, $^{56}$Fe and $^{208}$Pb which
we deem representative for the full mass range of stable nuclei.  We introduce
the relative and c.m.  coordinates and momenta
\begin{align}
\vec{r}_{12} &= { \vec{r} _1 - \vec{r} _2 }, \hspace{0.05\textwidth}
\vec{R}_{12} = \frac{\vec{r} _{1} + \vec{r} _{2}}{2} \; ,\\
\vec{k}_{12} &= \frac{ \vec{k} _1 - \vec{k} _2 } {2}, \hspace{0.05\textwidth}
\vec{P}_{12} = \vec{k} _{1} + \vec{k} _{2} \; . 
\end{align}
The corresponding two-body momentum density reads
\begin{multline}
   P_2 \left( \vec{k}_{12} , \vec{P}_{12} \right) = \frac{1}{ (2 \pi )^6} \int d
  \vec{r}_{12}
  \int d \vec{r}_{12}^{\;\prime} \int d \vec{R}_{12} \int d 
  \vec{R}_{12}^{\;\prime}
  \\ 
  \times e^{ \imath \vec{k}_{12} \cdot \left( \vec{r}_{12}^{\;\prime} -
\vec{r}_{12}
\right)} 
e^{ \imath \vec{P}_{12} \cdot \left( \vec{R}_{12}^{\;\prime} -
\vec{R}_{12}\right) }  
\rho_2 ( \vec{r}_{12}^{\;\prime}, \vec{R}_{12}^{\;\prime}; \vec{r}_{12},
\vec{R}_{12}),
\label{eq:defofp2}
\end{multline}
where $\rho_2 ( \vec{r}_{12}^{\;\prime}, \vec{R}_{12}^{\;\prime}; \vec{r}_{12} ,
\vec{R}_{12} )$ is
the non-diagonal 
two-body density (TBD) matrix 
\begin{multline}
   \rho_2 ( \vec{r}_{12}^{\;\prime}, \vec{R}_{12}^{\;\prime}; \vec{r}_{12},
\vec{R}_{12} ) 
    =  
  \int \{d \vec{r}_{3-A}\} 
   \\ 
    \times
\Psi_A^*(\vec{r}_1^{\;\prime},\vec{r}^{\;\prime}_2,\vec{r}_3,\ldots,\vec{r}_A) 
  \Psi_A(\vec{r}_1,\vec{r}_2,\vec{r}_3,\ldots,\vec{r}_A).
\end{multline}
Here, $\Psi_A$ is the normalized ground-state wave function of the
nucleus $A$ and $ \{ d\vec{r}_{i-A}\} \equiv
d\vec{r}_id\vec{r}_{i+1}\ldots d\vec{r}_A $.  For a spherically
symmetric system, $P_2 \left( \vec{k}_{12} , \vec{P}_{12} \right)$
depends on three independent variables, for example the magnitudes $
\left| \vec{k}_{12} \right|$ and $\left| \vec{P}_{12} \right|$ and the
angle between $\vec{k}_{12}$ and $\vec{P}_{12}$. In
Ref.~\cite{Alvioli:2012dd} two-body momentum distributions for $^3$He
and $^4$He are shown to be largely independent of the angle between
$\vec{k}_{12}$ and $\vec{P}_{12}$ for $P_{12} \lesssim $~200~MeV.
Integrating over the directional dependence of Eq.~(\ref{eq:defofp2}), 
the quantity  
\begin{align}
  & n_2( k_{12}, P_{12} ) k_{12}^2dk_{12} P_{12}^2 d P_{12} 
\nonumber \\ &
= k_{12}^2dk_{12} P_{12}^2 d P_{12} \int d \Omega_{k_{12}} \int d
\Omega_{P_{12}}  P_2 (\vec{k}_{12},
  \vec{P}_{12} ) \; , 
\label{eq:n2}
\end{align}
is connected to the probability of finding a nucleon pair with
relative and c.m. momentum in $[k_{12},k_{12}+dk_{12}]$ and
$[P_{12},P_{12}+dP_{12}]$.  With the spherical-wave expansion for the
two vector plane waves in Eq.~(\ref{eq:defofp2}) one obtains
\begin{equation}
  n_2( k_{12}, P_{12} )= \frac{4}{\pi^2} \sum_{l m_l } \sum_{ \Lambda M_\Lambda
} n_2^{l m_l \Lambda M_\Lambda}(k_{12},
  P_{12}) ,
  \label{eq:projection}
\end{equation}
with
\begin{align}
  &  n_2^{l m_l \Lambda M_\Lambda}(k_{12}, P_{12}) = \nonumber \\
   &  \int dr_{12}^{\;\prime}\; {r_{12}^{\;\prime}}^2 \int
   dR_{12}^{\;\prime}\; {R_{12}^{\;\prime}}^2 \int dr_{12} \; r_{12}^2 \int
   dR_{12}\; R_{12}^2 
   \nonumber \\ &  \times
   j_l(k_{12}r_{12}) j_l(k_{12}r_{12}^{\;\prime}) j_\Lambda(P_{12}R_{12})
j_\Lambda(P_{12}R_{12}^{\;\prime}) 
\nonumber \\
&   \times  \rho_2^{l m_l \Lambda
   M_\Lambda}(r_{12}^{\;\prime}, R_{12}^{\;\prime}; r_{12}, R_{12}).
   \label{eq:projected_n2}
\end{align}
Here, $\rho_2^{l m_l \Lambda M_\Lambda}(r_{12}^{\;\prime},
R_{12}^{\;\prime};r_{12}, R_{12})$ is the projection of the TBD matrix
on relative and c.m.~orbital angular-momentum states $\left| l m_l \right>$  and
$ \left| \Lambda M_\Lambda \right>$.

The pair c.m. momentum distribution is defined by
\begin{align}
  P_2 ( P_{12} )  & =   \int d \Omega_{P_{12}} \int d \vec{k}_{12} 
  P_2 (\vec{k}_{12}, \vec{P}_{12} ) \nonumber \\
  & = \int d k_{12} k_{12}^2 n_2(k_{12},P_{12}),
\label{eq:P2}
\end{align}
and the quantity $P_2(P_{12})~P_{12}^2~dP_{12}$ is related to the probability of
finding a nucleon pair with $\left| \vec{P}_{12} \right|$ in
$\left[ {P}_{12}, {P}_{12} + d {P}_{12} \right]$ irrespective of the
magnitude and direction of $ \vec{k}_{12}$. 
Similarly, the pair relative momentum distribution is defined as 
\begin{equation}
n_2(k_{12}) =  \int d \Omega_{k_{12}} \int d \vec{P}_{12} 
P_2 (\vec{k}_{12}, \vec{P}_{12} ) \; .
\label{eq:pairrelmomdis}
\end{equation}

In the IPM, the ground-state wave function can be expanded in terms of
single-particle wave functions $\phi_{\alpha_i}$  
\begin{equation}
  \Psi^\text{IPM}_A = (A!)^{-1/2} \mathrm{det}
\Bigl[ \phi_{\alpha_i}(\vec{x}_j) \Bigr] ,
  \label{eq:wf_mf}
\end{equation}
and the TBD matrix is given by
\begin{align}
  & 
  \rho^\text{IPM}_2 ( \vec{r}_{12}^{\;\prime}, \vec{R}_{12}^{\;\prime};
\vec{r}_{12}, \vec{R}_{12} )
   = \frac{2}{A(A-1)}
   \nonumber \\ &  \times
   \sum_{\alpha<\beta} 
   \frac{1}{2} 
   \left[ \phi^*_\alpha(\vec{x}_1^{\;\prime})
   \phi^*_\beta(\vec{x}_2^{\;\prime}) -
   \phi^*_\beta(\vec{x}_1^{\;\prime})
   \phi^*_\alpha(\vec{x}_2^{\;\prime}) \right] \nonumber \\ & \times
   \left[ \phi_\alpha(\vec{x}_1) \phi_\beta(\vec{x}_2) -
   \phi_\beta(\vec{x}_1) \phi_\alpha(\vec{x}_2) \right].
  \label{eq:tbd_mf}
\end{align}
Here, $\vec{x} \equiv \left( \vec{r}, \vec{\sigma}, \vec{\tau}
\right)$ is a shorthand notation for the spatial, spin, and isospin
coordinates. The summation $ \sum_{\alpha<\beta}$ extends over all
occupied single-particle levels and implicitly includes an integration
over the spin and isospin degrees of freedom (d.o.f.).

In a HO basis the uncoupled single-particle states read
\begin{equation}
  \phi_{\alpha}(\vec{x}) \equiv \psi_{n_\alpha l_\alpha m_{l_\alpha}}(\vec{r})
\chi _{\sigma_\alpha} \left( \vec{\sigma}\right) \xi_{\tau_\alpha} \left(
\vec{\tau} \right).
\label{eq:HOwf}
\end{equation}
The A dependence can be taken care of by means of the parameterization
$\hbar \omega (\textrm{MeV})=45 \; A ^ {− \frac{1}{3}} - 25
\;A^{−\frac{2}{3}}$.  A transformation from $(\vec{r}_1,\vec{r}_2)$ to
$(\vec{r}_{12}, \vec{R}_{12})$ for the uncoupled
normalized-and-antisymmetrized (nas) two-nucleon states can be readily
performed in a HO basis \cite{Vanhalst:2011es,Vanhalst:2012ur}
\begin{widetext}
\begin{align}
  \ket{\alpha\beta}_{\text{nas}}  = & 
  \sum_{\substack{nlm_lN\Lambda M_\Lambda \\ S M_S T M_T}} 
  \braket{nlm_l N\Lambda M_\Lambda  S M_S TM_T}{\alpha\beta}
  \ket{nlm_l N\Lambda M_\Lambda S M_S TM_T}
  = 
  \sum_{\substack{A=\{nlm_l N\Lambda M_\Lambda \\ S M_S T M_T\}}} 
  C^A_{\alpha\beta}
\ket{A} ,
  \label{eq:transformation}
\end{align}
with the transformation coefficient 
$C_{\alpha\beta}^{nl m_l N\Lambda M_\Lambda S M_S T M_T }$ 
given by
\begin{align}
C_{\alpha\beta}^{nl m_l N\Lambda M_\Lambda  S M_S T M_T }
  = & 
  \frac{1}{\sqrt{2}} \left[ 1 - (-1)^{l+S+T} \right] 
  \braket{\frac{1}{2} \tau_\alpha \frac{1}{2} \tau_\beta}{ T M_T} 
  \braket{\frac{1}{2} \sigma_\alpha \frac{1}{2} \sigma_\beta}{ S M_S} 
   \nonumber \\ & \times
  \sum_{L M_L}
  \braket{ l_\alpha m_{l_\alpha} l_\beta m_{l_\beta} }{L M_L}
  \braket{ nlN\Lambda;L}{ n_\alpha l_\alpha n_\beta l_\beta; L}_\text{SMB} 
  \braket{L M_L}{ l m_l \Lambda M_\Lambda}\,,
   \label{eq:transformationcoeff}
\end{align}
where we use the Talmi-Moshinsky brackets  $\langle
|\rangle_\text{SMB}$ \cite{moshinskyharmonic} to separate out the relative and
c.m. coordinates in the products of single-particle wave functions.

After performing the transformation of Eq.~(\ref{eq:transformation}) for the TBD
matrix of Eq.~(\ref{eq:tbd_mf}), $P_2(P_{12})$ 
can be written as
\begin{equation}
  P_2( P_{12} )= \frac{2}{\pi} \sum_{n l m_l } \sum_{ \Lambda M_\Lambda } P_2^{n
l m_l \Lambda M_\Lambda}(P_{12}) ,
  \label{eq:projectionP2}
\end{equation}
with 
\begin{multline}
   P_2 ^{n l m_l \Lambda M_\Lambda}(P_{12})= 
   \frac{2}{A(A-1)}
   \sum_{\alpha < \beta }
   \sum_{N N^\prime}
   \sum_{S M_S T M_T} 
   (C_{\alpha\beta}^{nlm_lN'\Lambda M_\Lambda S M_S T M_T })^\dagger 
   C_{\alpha\beta}^{nlm_lN\Lambda M_\Lambda S M_S T M_T } \\ \times
   \int dR_{12}^{\;\prime}\; {R_{12}^{\;\prime}}^2
   \int dR_{12} \; R_{12}^2 \;\;
   j_\Lambda(P_{12} R_{12}^{\;\prime})
   j_\Lambda(P_{12} R_{12})
   R_{N' \Lambda} ( \sqrt{2} R_{12}^{\;\prime} ) 
   R_{N \Lambda} ( \sqrt{2} R_{12} ) 
   \label{eq:projected_P2}
\end{multline}

\end{widetext}
A Woods-Saxon basis, for example, first needs to be expanded in a HO
basis before a projection of the type~(\ref{eq:projected_P2}) can be
made.  Using Eqs.~(\ref{eq:projectionP2}) and (\ref{eq:projected_P2}), 
the conditional pair c.m. momentum distribution for a given relative
radial quantum number $n$ and relative orbital momentum $l$, can be
defined as
\begin{equation}
   P_2  ( P_{12} | nl=\nu\lambda ) 
   = \frac{2}{\pi}  
  \sum_{m_l} \sum_{\Lambda M_\Lambda}  
  P_2^{\nu\lambda m_l \Lambda M_\Lambda}(P_{12}) \; .
  \label{eq:conditional_P2}
\end{equation}
Obviously, one has 
\begin{equation}
  P_2(P_{12}) 
  = \sum_{\nu \lambda} P_2(P_{12}|nl=\nu\lambda) 
  = \sum_{\lambda} P_2(P_{12}|l=\lambda) ,
\end{equation}
where $P_2(P_{12}|l=\lambda)$ is the conditional pair c.m. momentum distribution
for $l=\lambda$.

A symmetric correlation operator $\widehat{\cal{G}}$ can be applied to
the IPM wave function of Eq.~(\ref{eq:wf_mf}) in order to obtain a
realistic ground-state wave function
\cite{Pieper:1992gr,AriasdeSaavedra:2007qg,Engel:2011ss,Roth:2010bm}
\begin{equation}   
 \mid { \Psi_A}   \rangle =  \frac{1}
{ \sqrt{\langle \ \Psi  ^\text{IPM} _A \mid \widehat{\cal
G}^{\dagger} \widehat{\cal G} \mid \Psi  ^\text{IPM} _A \ \rangle}} \ 
\widehat
{ {\cal G}} \mid  \Psi  ^\text{IPM} _A \ \rangle \; .
\label{eq:realwf}
\end{equation}
The operator $\widehat{\cal{G}}$ is complicated but as
far as the SRC are concerned, it is dominated by
the central, tensor and spin-isospin correlations
\cite{janssen00,ryckebusch97}
\begin{equation}
\widehat{\mathcal{G}}   \approx   \widehat {{\cal S}}  
\biggl[ \prod _{i<j=1} ^{A} \biggl( 1 + \hat{\mathcal{o}} \left(\vec{x}_i,
\vec{x}_j \right) \biggr) \biggr]  \; ,
\label{eq:coroperator}
\end{equation}
with $ \widehat
{{\cal S}} $ the symmetrization operator and  
\begin{eqnarray}
\hat{\mathcal{o}} \left(\vec{x}_1, \vec{x}_2 \right)
& = & - g_c(r_{12}) + f_{t\tau}(r_{12}) S_{12} \vec{\tau}_1 \cdot 
  \vec{\tau}_2 \nonumber \\ & + & f_{\sigma\tau}(r_{12}) \vec{\sigma}_1 \cdot
\vec{\sigma}_2 
  \vec{\tau}_1 \cdot \vec{\tau}_2 \, ,
\label{eq:sumofcorrelators}
\end{eqnarray}
where $g_c(r_{12})$, $f_{t\tau}(r_{12})$, $f_{\sigma\tau}(r_{12})$ are the
central, tensor, and spin-isospin correlation functions, and ${S_{12}}$ the
tensor
operator.
The sign convention of $- g_c(r_{12})$ in Eq.~(\ref{eq:sumofcorrelators})
implies that
$ \displaystyle \lim _{r_{12} \to 0} g_c(r_{12})= g_0~(0<g_0 \le 1))$. 
We stress that the correlation functions cannot be considered as
universal \cite{Engel:2011ss}.
They depend for example on the
choices made with regard to the nucleon-nucleon
interaction, the single-particle basis and the many-body
approximation scheme.

\begin{figure}[tbp]
\includegraphics[width=\columnwidth]{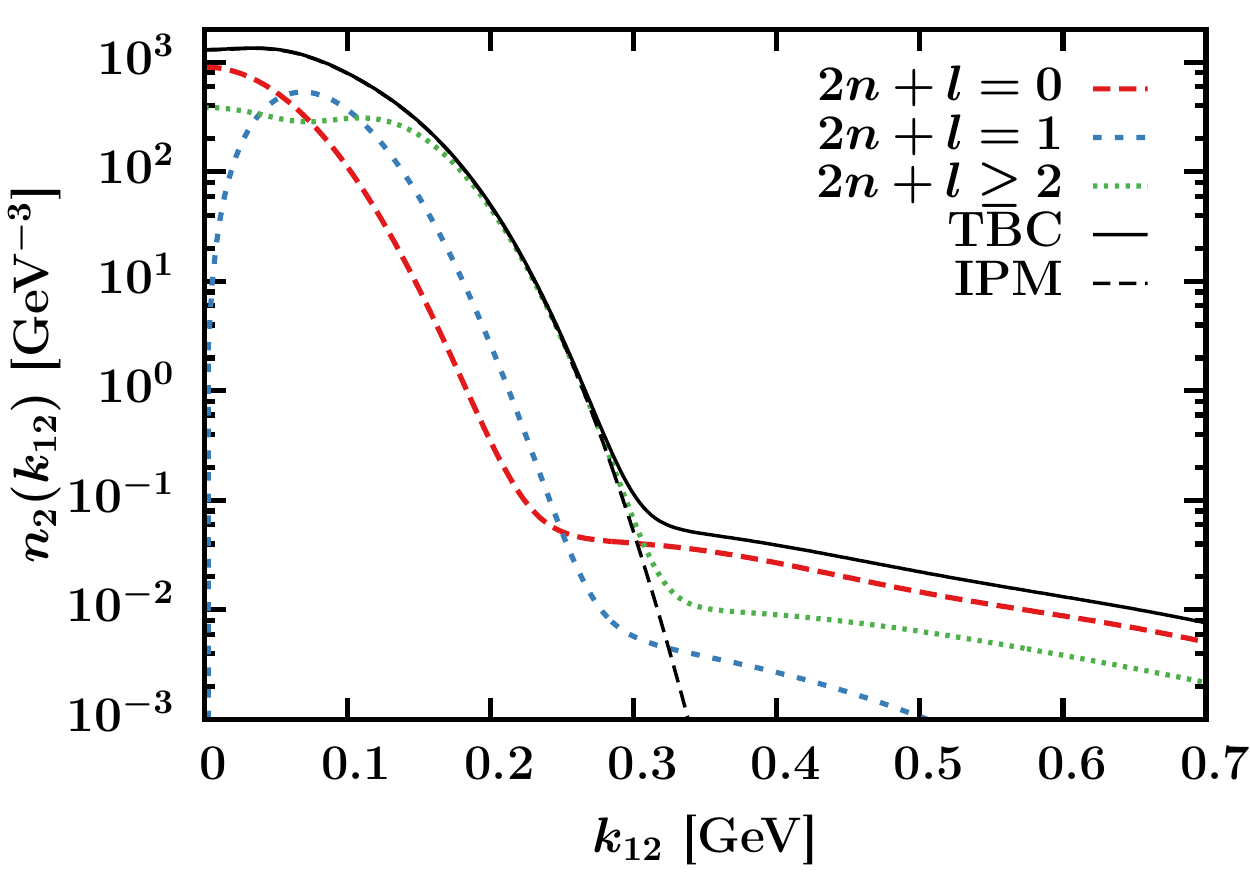}
\caption{(Color online) The momentum dependence of the computed
  $n_2^{2n+l}(k_{12})$, $n_2^\text{TBC}(k_{12})$ and $n_2^\text{IPM}(k_{12})$
  for $^{56}$Fe in a HO basis.  In order to
  quantify the effect of SRC we have used the $g_c \left( r_{12}
  \right)$ of Ref.~\cite{gearheart94} and the $f_{t\tau}\left( r_{12}
  \right)$, $f_{\sigma\tau}(r_{12})$ of Ref.~\cite{Pieper:1992gr}.}
\label{fig:corrcontrib}
\end{figure}

With Eq.~(\ref{eq:realwf}), the intrinsic complexity stemming from
the nuclear correlations is shifted from the wave functions to the
transition operators. For example, the ground-state matrix element
with a two-body operator $\hat{\mathcal{O}}^{[2]}$ adopts the form
\begin{eqnarray}
\langle { \Psi_A} \mid \hat{\mathcal{O}}^{[2]} \mid { \Psi_A}   \rangle 
& = &  \frac{1}
{ {\langle \ \Psi  ^\text{IPM} _A \mid \widehat{\cal G}^{\dagger} \widehat{\cal
G} \mid \Psi  ^\text{IPM} _A \ \rangle}} 
\nonumber \\
& \times & 
\langle \ \Psi  ^\text{IPM} _A \mid \widehat{\cal G}^{\dagger}  
\hat{\mathcal{O}}^{[2]} 
\widehat
{ {\cal G}} \mid  \Psi  ^\text{IPM} _A \ \rangle \; ,
\label{eq:transgen}
\end{eqnarray} 
whereby high-order many-body operators are generated.
Throughout this work we adopt the two-body cluster (TBC) approximation,
which amounts to discarding all terms in $\widehat{\cal G}^{\dagger}
\hat{\mathcal{O}}^{[2]} \widehat { {\cal G}} $ except those in which
the transition operator and the correlators act on the same pair of
particles. In this lowest-order cluster expansion the matrix element  of
Eq.~(\ref{eq:transgen}) becomes with the aid of Eq.~(\ref{eq:coroperator})
\begin{eqnarray}
& & \langle { \Psi_A} \mid \hat{\mathcal{O}}^{[2]} \mid { \Psi_A}   \rangle 
 \approx  \frac{1}
{ {\langle \ \Psi  _A \mid   \Psi  _A \ \rangle}} 
\nonumber \\
&& \times \langle \ \Psi  ^\text{IPM} _A \mid
\sum _{i<j=1} ^{A}
  \biggl( 1 + \hat{\mathcal{o}} \left(\vec{x}_i, \vec{x}_j \right) \biggr)  ^
{\dagger}
\hat{\mathcal{O}}^{[2]} \left( i, j \right)
\nonumber \\
& & \times 
 \biggl( 1 + \hat{\mathcal{o}} \left(\vec{x}_i, \vec{x}_j \right) \biggr)
\mid \Psi  ^\text{IPM} _A \rangle
\nonumber \\
& & = \frac{1}
{ \langle  \Psi  _A \mid   \Psi  _A \ \rangle} 
\nonumber \\
& & \times \left[ \langle \ \Psi  ^\text{IPM} _A \mid   
\hat{\mathcal{O}}^{[2]} 
\mid  \Psi  ^\text{IPM} _A \ \rangle + \textrm{TBC corrections} \right]
   . 
\label{eq:translowestorder}
\end{eqnarray}
In this expansion, the matrix element is written as the sum of the
bare (or IPM) contribution and the TBC corrections to it. The $P_{2}
(P_{12} )$ and $n_{2} (k_{12})$ of
Eqs.~(\ref{eq:P2}-\ref{eq:pairrelmomdis}) can be computed with the aid
of the Eq.~(\ref{eq:translowestorder}) using the transition operators
$\delta \left( \vec{P}_{ij} - (\vec{k}_{i} + \vec{k} _j) \right)$ and
$\delta \left( \vec{k}_{ij} - \frac {\vec{k}_{i} - \vec{k} _j} {2}
\right)$. As the $\hat{\mathcal{o}}$ involves only relative
coordinates, the $P_{2}(P_{12} )$ is not affected by the SRC
corrections in the TBC approximation.  We define $n_2^\text{IPM} (k_{12})$
as the IPM contribution of $n_2 (k_{12})$ and $n_2^\text{TBC} (k_{12})$ the
result obtained with Eq.~(\ref{eq:translowestorder}). Accordingly,
$n_2^\text{TBC}(k_{12}) = n_2^\text{IPM}(k_{12}) + \textrm{TBC corrections}$. 
For
$n_{2} ^\text{TBC} (k_{12})$ the denominator ${ {\langle \ \Psi _A \mid
    \Psi _A \ \rangle}} $ in Eq.~(\ref{eq:translowestorder}) can be
numerically computed by imposing the normalization conditions: $\int d
k_{12} n_{2}^\text{TBC} (k_{12}) k_{12}^{2} = 1$.  As in
Eqs.~(\ref{eq:projected_n2}) and (\ref{eq:conditional_P2}), one can
introduce projection operators, and select the contributions to $n_{2}
^\text{TBC} (k_{12})$ stemming from particular quantum numbers $(nl)$ of
the relative two-nucleon wave functions in $\Psi ^\text{IPM} _A$. We define
$n_2^{2n+l}(k_{12})$ as the contribution to $n_2^\text{TBC}$ considering
only $(nl)$ configurations in $\Psi ^\text{IPM} _A$ with constant $2n+l$.
Obviously, one has
\begin{equation}
  \sum_{2n+l} n_2^{2n+l}(k_{12}) = n_2^\text{TBC}(k_{12}).
\end{equation}
The computed $n_2^{2n+l}$, $n_2^\text{TBC}$ and $n_2^\text{IPM}$ for $^{56}$Fe
are shown in
Fig.~\ref{fig:corrcontrib}. Below the Fermi momentum $k_F$, the effect
of the correlation operator is negligible and $n_2^\text{IPM}(k_{12})
\approx n_2 ^\text{TBC}(k_{12})$.  For $k_{12}> k_F$, $n_2^\text{IPM}(k_{12})$
drops rapidly while $n_2 ^\text{TBC}(k_{12})$ exhibits the SRC related high
momentum tail. The tail is dominated by the $2n+l=0$ configurations. This
indicates that most of the SRC are dynamically generated through the
operation of the correlation operators on $nl=00$ IPM pairs.

\begin{figure}[htbp]
  \includegraphics[width=\columnwidth]{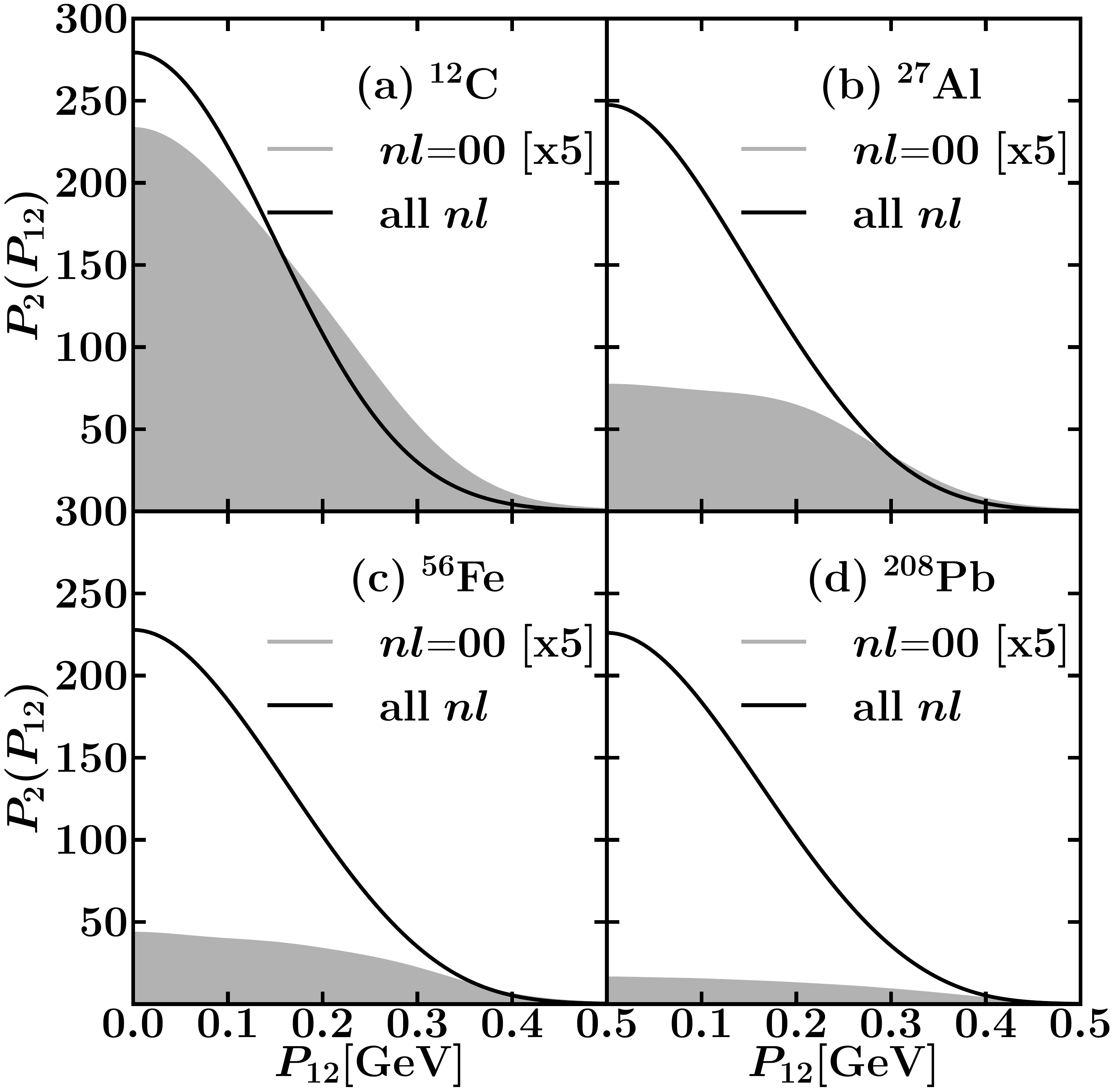}
  \caption{ The momentum dependence of $P_2(P_{12})$ and the
    $P_2\left( P_{12} | nl =00 \right)$ for pp pairs in different
    nuclei.  The adopted normalization convention is that
    $\int_0^\infty \mathrm{d}P_{12}\; P_{12}^2 P_2(P_{12}) = 1$. Note
    that only the pp contributions to $P_2(P_{12})$ are considered
    when performing the integral. The results are obtained in a HO
    basis. 
  }
  \label{fig:mompp_cm}
\end{figure}
%
\begin{figure}[htbp]
  \includegraphics[width=\columnwidth]{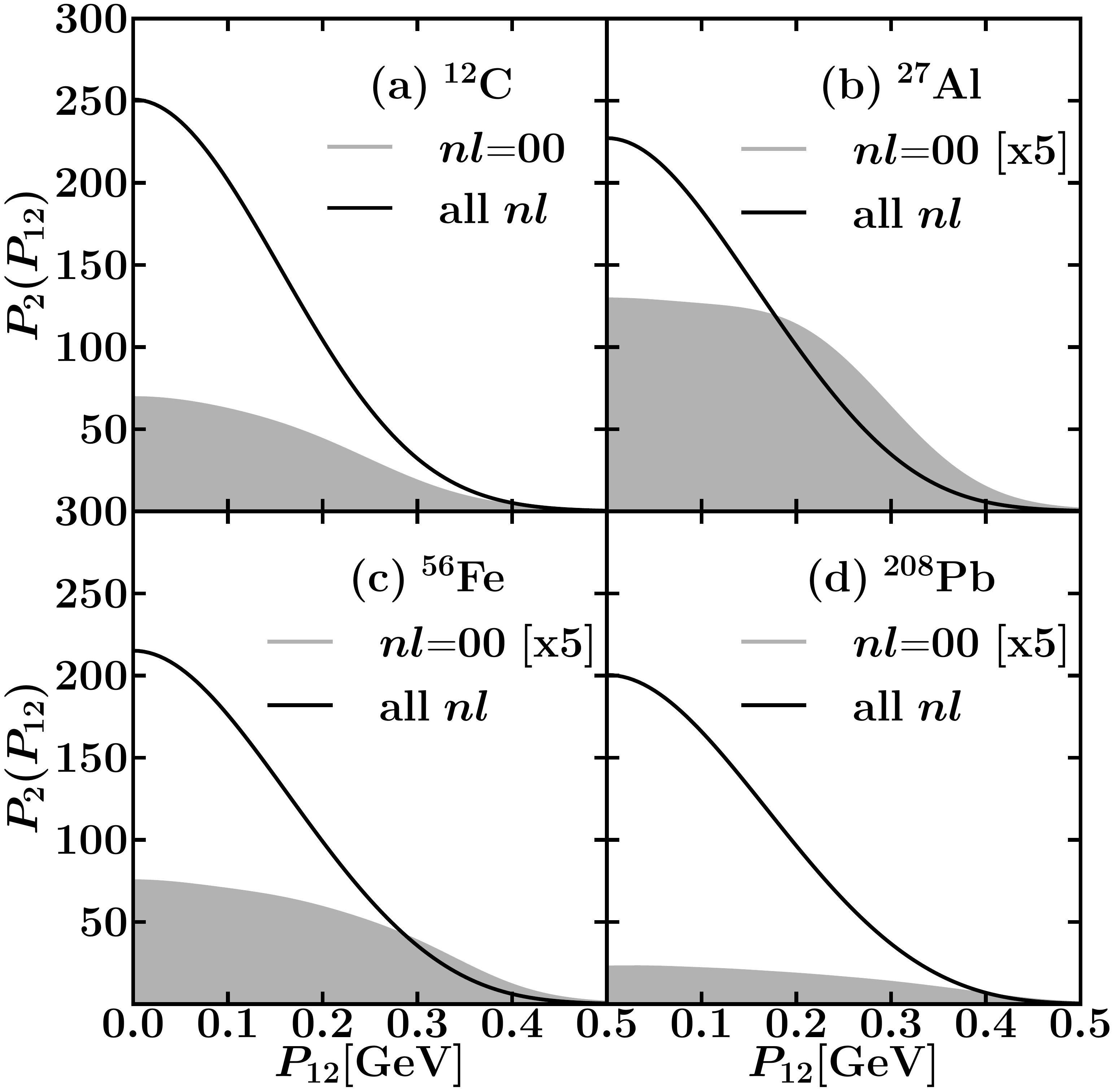}
  \caption{ As in Fig.~\ref{fig:mompp_cm} but for pn pairs.}
  \label{fig:mompn_cm}
\end{figure}

In Sec.~\ref{sec:fac}, it is shown that in the limit of vanishing FSIs
the factorization function of the exclusive $A(e,e'pN)$ cross section is
$P_{2} (P_{12}|nl=00) $.  In Figs.~\ref{fig:mompp_cm}
and~\ref{fig:mompn_cm}, we display the computed $P_2(P_{12})$ and
$P_2(P_{12}|nl=00)$ for the pp and pn pairs in $^{12}$C, $^{27}$Al,
$^{56}$Fe and $^{208}$Pb.  The relative weight of the $(nl=00)$ in the
total c.m.~distribution decreases spectacularly with increasing mass
number $A$.  This will reflect itself in the mass dependence of the
$A(e,e'NN)$ cross sections which are predicted to scale much softer than
$A^2$.  The $(nl=00)$ pairs are strongly localized in space which
enlarges the $P_2(P_{12}|nl=00)$ width relative to the 
$P_2(P_{12})$ one. The mass dependence of the normalized $P_2(P_{12})$
reflects itself in a modest growth of the width of the distribution.
For the light nuclei $^{12}$C and $^{27}$Al, the pp and pn
c.m. distributions look very similar.

At first sight the computed $P_2(P_{12})$ for the pp and pn pairs in
Figs.~\ref{fig:mompp_cm} and \ref{fig:mompn_cm} look very Gaussian.
In what follows, we use the moments to quantify the non-Gaussianity of
the $P_2$.  The first moment, or mean, of a distribution $F(x)$ is
defined as
\begin{equation}
  \mu_1 = \mu = \frac{\int_{D} x F(x) \mathrm{d} x}
  { \int_{D} F(x) \mathrm{d}x } ,
  \label{eq:moment1}
\end{equation}
where $D$ is the domain of the distribution.  For 
$m>1$, we define the central moments as
\begin{equation}
  \mu_m = \frac{\int_{D} (x-\mu)^m F(x)  \mathrm{d} x}
  { \int_{D} F(x) \mathrm{d}x } \; .
\end{equation}
The width is defined as $\sigma = \sqrt{\mu_2}$.  With regard to
$\mu_3$ and $\mu_4$, it is common practice to describe a distribution
with the skewness $\gamma_1$ and excess kurtosis $\kappa$
\begin{eqnarray}
  \gamma_1 & \equiv & \frac{\mu_3}{\sigma^3} \label{eq:skewness} \\
  \kappa & \equiv & \frac{\mu_4}{\sigma^4} - 3 \label{eq:kurtosis}, 
\end{eqnarray}
which are both vanishing for a Gaussian distribution. 

For a spherically symmetric distribution, one can derive the
distributions $P_{2,i} \left(P_{12,i} \right) \; (i=x,y,z)$ along the
axes from $ P_{12}^2 P_2 \left( P_{12} = \sqrt{P_{12,x}^2 + P_{12,y}^2
  + P_{12,z}^2} \right) $.  Gaussian $ P_{2,i} $ give rise to a
$P_{12}^2 P_2 \left(P_{12,i} \right) $ of the Maxwell-Boltzmann type.

\begin{figure*}[htbp]
  \centering \includegraphics[viewport=0 0 320 225 , clip,
    width=0.75\textwidth] {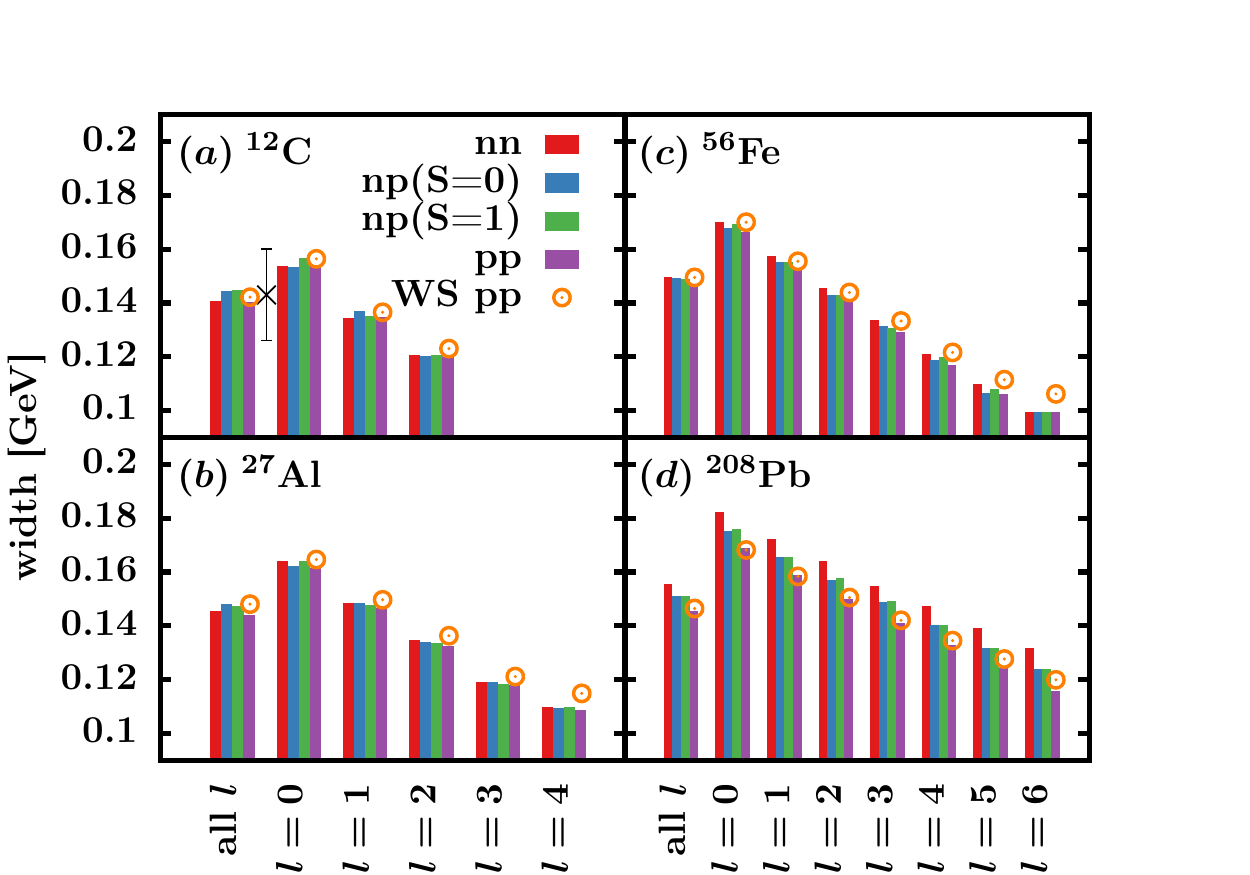}
  \caption{(Color online) Computed widths of the
    $P_{2,x}\left(P_{12,x}\right)$ (denoted as ``all $l$'') and
    $P_{2,x} \left(P_{12,x}|l\right)$ distributions for pp, nn,
    np$(S=0)$ and np$(S=1)$ pairs in $^{12}$C, $^{27}$Al, $^{56}$Fe,
    $^{208}$Pb. Unless stated otherwise the results are obtained in a
    HO basis.  For pp pairs we also display results for a WS basis
    (denoted as ``WS pp'').  The black cross is the experimental result
    from Ref.~\cite{Tang:2002ww}.  }
  \label{fig:moments_all}
\end{figure*}

\begin{table*}[htbp]
  \centering
  \begin{tabular}{ccccccccc}
   \hline \hline 
    & & \multicolumn{3}{c}{HO} & &
    \multicolumn{3}{c}{WS} \\
    \cline{3-5} \cline{7-9} & & & & & & & & \\    
    &   
    & $\sigma$ (MeV)& $\gamma_1$ [Eq.~(\ref{eq:skewness})]  
    & $\kappa$ [Eq.~(\ref{eq:kurtosis})] &  \phantom{qqqq}
    & $\sigma$ (MeV)& $\gamma_1$ [Eq.~(\ref{eq:skewness})]  
    & $\kappa$ [Eq.~(\ref{eq:kurtosis})] \\
    \hline 
    $^{12}$C & $P_{2,x}(P_{12,x}|nl=00)$ &  $ 156$ & $0.00$  & $-0.25$ &
    &  $ 158$ & $0.00$  & $-0.28$\\
    $^{12}$C & $P_{2,x}(P_{12,x})$ &  $140$ & $-0.01$  & $-0.12$ &
    & $142$ & $-0.01$ & $-0.05$ \\
    $^{27}$Al & $P_{2,x}(P_{12,x}|nl=00)$ &  $164$ & $0.00$ & $-0.45$ & 
    &  $168$ & $0.00$ & $-0.45$ \\
    $^{27}$Al & $P_{2,x}(P_{12,x})$ &  $144$ & $-0.01$ & $-0.20$ &
    & $148$ & $-0.01$ & $-0.20$ \\
    $^{56}$Fe & $P_{2,x}(P_{12,x}|nl=00)$& $172$ & $0.00$  & $-0.54$ & 
    & $174$ & $0.00$  & $-0.54$ \\
    $^{56}$Fe & $P_{2,x}(P_x)$& $146$ & $-0.01$  & $-0.27$ &
    & $149$ & $0.00$ & $-0.26$ \\
    $^{208}$Pb & $P_{2,x}(P_{12,x}|nl=00)$& $178$ & $0.00$  & $-0.58$ & 
    & $177$ & $0.00$  & $-0.63$ \\
    $^{208}$Pb & $P_{2,x}(P_{12,x})$& $145$ & $0.00$  & $-0.31$ &
    & $146$ & $0.00$ & $-0.31$ \\
    \hline \hline
  \end{tabular}
  \caption{ The moments of the $P_{2,x} \left (P_{12,x} \right)$ and
    the $P_{2,x} \left( P_{12,x}|nl=00 \right)$ distributions for pp
    pairs as computed in a HO and WS single-particle basis for various
    nuclei.}
  \label{tab:moments}
\end{table*}

Table~\ref{tab:moments} shows the computed moments of the
$P_{2,x}(P_{12,x}|nl=00)$ and $P_{2,x}(P_{12,x})$ distributions for pp
pairs.  These results are obtained with HO and Woods-Saxon (WS)
single-particle wave functions. We find that the c.m. distributions
are not perfectly Gaussian and that the non-Gaussianity grows with
$A$. The values of the widths are only moderately sensitive to the
single-particle basis used. The WS widths are larger by a few percent
than the HO ones.

In Fig.~\ref{fig:moments_all}, the calculated widths of the
$P_{2,x}(P_{12,x})$ and $P_{2,x}(P_{12,x}|l)$ are shown for pp, nn and
np pairs. For the np pairs we discriminate between singlet $(S=0)$ and triplet
$(S=1)$ spin states.  From Fig.~\ref{fig:moments_all} we draw the following
conclusions.  The width of the $P_{2,x}(P_{12}|l)$ depends on $l$.
For $l=0$ and np pairs, the width of $P_{2,x}(P|l)$ is almost
independent of $S$.  For heavy nuclei there is a substantial
difference in the width of the $P_{2,x}(P|l=0)$ for pp, nn and np
pairs but for light nuclei this is not the case.  
A similar but smaller dependence on the width is found for $n$ at fixed $l$,
the width of $P_2(P_{12}|nl)$ decreases for increasing $n$.
We conclude that
from the width of the c.m. distribution of the pairs one can infer
information about their relative orbital momentum.

\section{Factorization of the two-nucleon knockout cross section} \label{sec:fac}
It is well known that the fivefold differential cross section for the
exclusive $A(e, e' p)A-1$ reaction under quasifree kinematics with
$A-1$ spectators
\begin{eqnarray}
\label{eq:eepreaction}
&&\gamma ^{*} \left( q \right) +  A - 1  \left( p_{A-1} \right) 
+ N \left( k_{1} \right)  \longrightarrow
\nonumber \\
&&A - 1  \left( p_{A-1} \right) + N \left( p_{1} \right) \;  , 
\end{eqnarray}
factorizes as  
\begin{equation}
d ^{5} \sigma (e,e'p) = K_{ep} \sigma _{ep} P_1(\vec{k}_m,E_m) \; .
\label{eq:faceep}
\end{equation}
Here, $K_{ep}$ is a kinematical factor and $\sigma _{ep}$ the
off-shell electron-proton cross section. Further,
$\vec{k}_m=-\vec{p}_{A-1} = \vec{k}_1$ is the missing momentum and $E_{m}=q^0 -
T_{p_{1}} - T_{A-1}$ the missing energy, whereby $T_{A-1}$ and
$T_{p_{1}} $ are the kinetic energy of the recoiling nucleus and
ejected nucleon.  The $P_1(\vec{k},E)$ is the one-body spectral function and is
associated with the combined probability of removing a proton with
momentum $\vec{k}$ from the ground-state of $A$ and of finding the
residual $A-1$ nucleus at excitation energy $E$ (measured relative to
the ground-state of the target nucleus).  The factorization is exact in a
non-relativistic reaction model with $A-1$ spectators and vanishing
FSIs \cite{Caballero:1997gc}. The validity of the spectator
approximation requires that the $E_{m}$ is confined to low values,
corresponding to states with a predominant one-hole character relative
to the ground state of the target nucleus $A$.

Below, it is shown that also the $A(e,e'pN)$ differential cross section
factorizes under certain assumptions. The factorization function is
connected to the c.m.~motion of close-proximity pairs.  In
Ref.~\cite{Frankfurt:1988nt} the factorization function is introduced as the
so-called \textit{decay function}.  In Ref.~\cite{Ryckebusch:1996wc} a
factorized expression for the $A(e,e'pp)$ cross section has been
derived. Thereby, in computing the matrix elements, all FSI effects
have been neglected and the zero-range approximation $( \lim _{r_{12}
  \to 0})$ has been adopted.  A $^{12}$C$(e,e'pp)$ experiment conducted
at the Mainz Microtron (MAMI) \cite{Blomqvist:1998gq} showed very good
quantitative agreement with the predicted diproton pair c.m. momentum
factorization up to momenta of about 500~MeV. Here, the formalism of
Ref.~\cite{Ryckebusch:1996wc} is extended to include the effect of FSIs
and to soften the zero-range approximation. Note that the limit $ \lim
_{r_{12} \to 0}$ effectively amounts to projecting on states with
vanishing relative orbital momentum.

We consider exclusive $A(e,e'NN)$ reactions in the spectator
approximation with a virtual photon coupling to a correlated pair $N
\left( k_{1} \right) N \left( k_{2} \right)$
\begin{eqnarray}
\label{eq:eepNreaction}
&&\gamma ^{*} \left( q \right) +  A - 2  \left( p_{A-2} \right) 
+ N \left( k_{1} \right) N \left( k_{2} \right) \longrightarrow
\nonumber \\
&&A - 2  \left( p_{A-2} \right) + N \left( p_{1} \right) + N \left( p_{2}
\right) \; . 
\end{eqnarray}
In a non-relativistic treatment, the corresponding matrix element is given by
\begin{align} 
\label{eq:matrixelstart}
 \mathcal{M}^\mu  =&  \int d\vec{x}_1 \int d\vec{x}_2 \Big[
\chi^\dagger_{s_1}(\vec{\sigma}_1)  \xi^{\dagger}_{t_1} \left( \vec{\tau}_1 \right) 
\chi^\dagger_{s_2}(\vec{\sigma}_2)
 \xi^{\dagger}_{t_2} \left( \vec{\tau}_2 \right) \nonumber\\
&\times e^{-i\vec{p}_1\cdot\vec{r}_1}
e^{-i\vec{p}_2\cdot\vec{r}_2}  - (1
\leftrightarrow 2) \Big] \nonumber\\
&\times
\mathcal{F}^\dagger_{\text{FSI}}(\vec{r}_1,\vec{r}_2)
\hat{\mathcal{O}}^\mu(\vec{x}_1,\vec{x}_2)
\phi_{\alpha_1}(\vec{x}_1)\phi_{\alpha_2}(\vec{x}_2)\,.
\end{align}

Here, $s_i (t_i)$ are the spin (isospin) projection of the outgoing
nucleons. Further, $\mathcal{F}_\text{FSI}(\vec{r}_1,\vec{r}_2)$ is an
operator encoding the FSIs for a reaction where two nucleons are
brought into the continuum at the spatial localizations $\vec{r}_1$
and $\vec{r}_2$ respectively. We assume that $\mathcal{F}_\text{FSI}$
does not depend on the spin and isospin d.o.f, which is a fair
approximation at higher energies.  The amplitude of
Eq.~(\ref{eq:matrixelstart}) refers to the physical situation whereby,
as a result of virtual-photon excitation, two nucleons are excited
from bound states $\alpha_{1} \alpha_{2}$ into continuum states.

In Eq.~(\ref{eq:matrixelstart}), the effect of the correlations is
implemented in the TBC approximation by means of a
symmetric two-body operator \cite{Engel:2011ss,janssen00}
\begin{multline}
\hat{\mathcal{O}}^\mu(\vec{x}_1,\vec{x}_2)  = \\
 \biggl[ e^{i\vec{q}\cdot\vec{r}_1} 
\Gamma^\mu_{\gamma ^{\star} N} (\vec{x}_1) + 
e^{i\vec{q}\cdot\vec{r}_2} \Gamma^\mu_{\gamma ^{\star} N}  (\vec{x}_2)
\biggr] 
\hat{\mathcal{o}} \left(\vec{x}_1, \vec{x}_2 \right)\,,
\label{eq:twobodphotcoupling}
\end{multline}
where the operator $\hat{\mathcal{o}} \left(\vec{x}_1, \vec{x}_2
\right)$ has been defined in Eq.~(\ref{eq:sumofcorrelators}) and
$\vec{q}$ is the three-momentum of the virtual photon. The
$\Gamma^\mu_{\gamma ^{\star} N} (\vec{x}_i)$ denotes the one-body
virtual photon coupling to a bound nucleon with coordinate $\vec{x}_i$
(includes the spatial, spin, and isospin d.o.f.). The
Eq.~(\ref{eq:twobodphotcoupling}) can be interpreted as the
SRC-corrected photo-nucleon coupling which operates on IPM many-body
wave functions.

\begin{figure*}
  \centering
  \includegraphics[width=\textwidth]{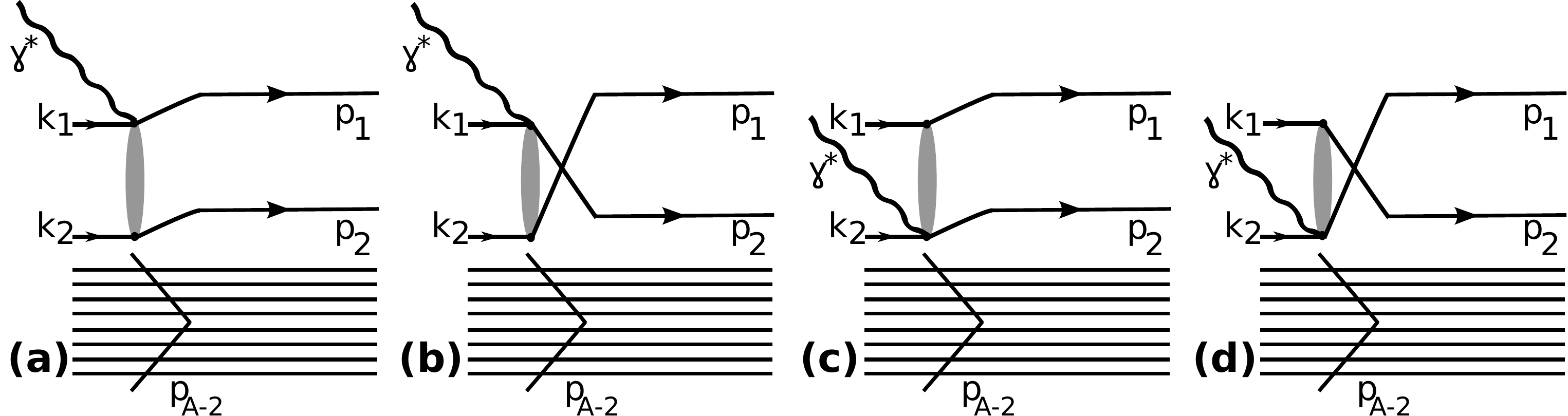}
  \caption{The four contributions to the $A(e,e'NN)$ amplitude of Eq.~(\ref{eq:matrixelstart}).
  }
  \label{fig:scheme1}
\end{figure*}

The amplitude of Eq.~(\ref{eq:matrixelstart}) involves four
contributions schematically shown in Fig.~\ref{fig:scheme1}.  For the
sake of brevity, in the following we consider the term of
Fig.~\ref{fig:scheme1}(a) with a photon-nucleon coupling on coordinate
$\vec{r}_1$ and the outgoing nucleon with momentum $\vec{p}_1$
directly attached to this vertex.  The corresponding amplitude is
denoted by $\mathcal{M}^\mu_{a}$.  The other three terms in
Fig.~\ref{fig:scheme1} follow a similar derivation.

In a HO single-particle basis, one can write 
\begin{align} \label{eq:matrixHOized}
 \mathcal{M}^\mu_{a}&
=\int d\vec{r}_1 \int
d\vec{r}_2 e^{-i(\vec{p}_1-\vec{q})\cdot\vec{r}_1}
e^{-i\vec{p}_2\cdot\vec{r}_2} 
\mathcal{F}^\dagger_{\text{FSI}}(\vec{r}_1,\vec{r}_2)
\nonumber\\\times&
\bra{s_1t_1,s_2t_2}
\Gamma^\mu_{\gamma ^{\star} N} \left(\vec{x}_1 \right)
\hat{\mathcal{o}}\left(\vec{x}_1, \vec{x}_2 \right) 
\ket{\sigma_1\tau_1,\sigma_2\tau_2}
\nonumber\\
\times&\psi_{n_1l_1m_{l_1}}(\vec{r}_1)\psi_{n_2l_2m_{l_2}}(\vec{r}_2)\, ,
\end{align}
where $\sigma_i$ $(\tau _i)$ are the spin (isospin) quantum numbers of
the bound states. Further, $\psi_{n_{1}l_{1}m_{l_{1}}}$ and
$\psi_{n_{2}l_{2}m_{l_{2}}}$ are the radial HO wave functions as
introduced in Eq.~(\ref{eq:HOwf}).

Similar to the Eq.~(\ref{eq:transformation}), we apply the
Talmi-Moshinsky brackets $\langle |\rangle_\text{SMB}$
\cite{moshinskyharmonic} to transform Eq.~(\ref{eq:matrixHOized}) to
relative and c.m. radial coordinates to obtain
\begin{widetext}
\begin{align}
\label{eq:relcmampl}
&\mathcal{M}^\mu_{a}  =
\sum_{\substack{L M_L}} \sum_{\substack{n l
m_l\\N \Lambda M_\Lambda}} \int d\vec{r}_{12} \int
d\vec{R}_{12}e^{-i\vec{P}_{12} \cdot\vec{R}_{12}} e^
{-i\vec{k}^-\cdot\vec{r}_{12}}
\mathcal{F}^\dagger_{\text{FSI}}(\vec{R}_{12} + 
\frac{\vec{r}_{12}}{2},\vec{R}_{12}-\frac{
  \vec { r }_{12}}{2})
\psi_{nlm_l}(\frac{\vec { r
} _ { 12 } } { \sqrt { 2}})
\psi_{N\Lambda M_\Lambda}(\sqrt{2}\vec{R}_{12}) 
\nonumber \\ & \times
\braket{l_1m_{l_1}l_2m_{l_2}}{LM_L} \braket{lm_l\Lambda
M_\Lambda}{LM_L}
\braket{nlN\Lambda ;L}{n_1l_1n_2l_2;L}_{\text{SMB}}
\bra{s_1t_1,s_2t_2}
\Gamma^\mu_{\gamma ^{\star} N} \left(\vec{x}_1 \right) 
\hat{\mathcal{o}} \left(\vec{x}_1, \vec{x}_2 \right)
\ket{\sigma_1\tau_1,\sigma_2\tau_2} \; ,
\end{align}
\end{widetext}
where $\vec{P}_{12} = \vec{p}_1 + \vec{p}_2 - \vec{q}$, $\vec{k}^\mp =
\frac{\vec{p}_1-\vec{p}_2}{2} \mp \frac{\vec{q}}{2}$. 

In Eq.~(\ref{eq:relcmampl}) the sum over the relative quantum numbers
is dominated by $(nl=00)$.  This is based on the observation that
typical correlation operators act over relatively short internucleon
distances and mostly affect the $(nl=00)$ components of the
$\psi_{nlm_l}$ wave functions. For a more detailed explanation we
refer to the discussion of Fig.~\ref{fig:corrcontrib} in
Sect.~\ref{sec:momdistr} and
Refs.~\cite{Vanhalst:2011es,Vanhalst:2012ur}.

For close-proximity nucleons one can set
$\vec{r}_{12}\approx \vec{0}$ in the FSI operator:
\begin{eqnarray} 
\mathcal{F}_{\text{FSI}}(\vec{r}_{1},\vec{r}_{2}) & = &
\mathcal{F}_{\text{FSI}}(\vec{R}_{12}+\frac{\vec{r}_{12}}{2},\vec{R}_{12}-\frac{
\vec { r }_{12}}{2}) 
\nonumber \\ &\approx&
\mathcal{F}_{\text{FSI}}(\vec{R}_{12},\vec{R}_{12})\,.\label{eq:fsiapprox}
\end{eqnarray}
This approximation amounts to computing the effect of FSIs as if the
the two nucleons are brought into the continuum at the same spatial
point (determined by the c.m. coordinate of the pair), which is very
reasonable for close-proximity nucleons.  With the above assumptions
one arrives at the expression for the matrix element
\begin{align}
\label{eq:relcmampl00}
\mathcal{M}^\mu_{a}&  \approx
\bra{s_1t_1,s_2t_2} \widehat{\Gamma}^\mu_{\gamma ^{\star} N}(\vec{k}^-) 
\ket{\sigma_1\tau_1,\sigma_2\tau_2}
\nonumber\\\times&
\sum_{\substack{N \Lambda M_\Lambda}}  
\braket{l_1m_{l_1}l_2m_{l_2}}{\Lambda M_\Lambda}
\braket{00N\Lambda ;\Lambda}{n_1l_1n_2l_2 ;\Lambda}_{\text{SMB}}
\nonumber \\ \times &
\int
d\vec{R}_{12}e^{-i\vec{P}_{12}\cdot\vec{R}_{12}}
\mathcal{F}^\dagger_ { \text {FSI}}(\vec{R}_{12},\vec{R}_{12})
\psi_{N\Lambda M_\Lambda}(\sqrt{2}\vec{R}_{12}) \,,
\end{align}
with
\begin{align}
 \widehat{\Gamma}^\mu_{\gamma ^{\star} N}(\vec{p}) \equiv \int d\vec{r}_{12}e^
{-i\vec{p}\cdot\vec{r}_{12}}  \psi_{000}(\frac{\vec { r
} _ { 12 } } { \sqrt { 2}})
\Gamma^\mu_{\gamma ^{\star} N} (\vec{x}_1)
\hat{\mathcal{o}}(\vec{x}_1, \vec{x}_2) \,.
\end{align}
In deriving the Eq.~(\ref{eq:relcmampl00}), we have separated the
integration over the spatial and spin-isospin d.o.f.. In addition, use
has been made of the fact that the operator
$\hat{\mathcal{o}}(\vec{x}_1, \vec{x}_2) $ of
Eq.~(\ref{eq:sumofcorrelators}) does not depend on the c.m. coordinate
$\vec{R}_{12}$.  The most striking feature of
Eq.~(\ref{eq:relcmampl00}) is the factorization of the amplitude
in a
term connected to the c.m. motion of the initial pair and a term which
contains the full complexity of the photon-nucleon coupling to a
correlated pair. 

After summing the four terms that contribute to Eq.~(\ref{eq:matrixelstart})
and squaring the matrix element, the eightfold differential cross section 
factorizes according to 
\begin{equation}
d ^{8} \sigma (e,e'NN) = K_{eNN} \sigma _{e2N} F^D_{n_1l_1,n_2l_2}(\vec{P}_{12}),
\label{eq:eeNNfactorized}
\end{equation}
with $K_{eNN}$ a kinematic factor. 
Further,  the off-shell
electron-two-nucleon cross section is given by
\begin{equation}
 \sigma_{e2N} \propto
L_{\mu\nu} \sum_{\substack{s_1s_2\sigma_1\sigma_2\\\tau_1\tau_2}}
J^\mu \left(J^{\nu }\right)^\dagger\,,
\end{equation}
with $L_{\mu\nu}$ the leptonic tensor and $J^\mu$ the hadronic current 
given by
\begin{align}
 J^\mu &= \bra{s_1t_1,s_2t_2} \widehat{\Gamma}^\mu_{\gamma ^{\star}
N}(\vec{k}^-) 
\ket{\sigma_1\tau_1,\sigma_2\tau_2}\nonumber\\&
-\bra{s_2t_2,s_1t_1} \widehat{\Gamma}^\mu_{\gamma ^{\star} N}(\vec{k}^+) 
\ket{\sigma_1\tau_1,\sigma_2\tau_2}\nonumber\\&
+\bra{s_1t_1,s_2t_2} \widehat{\Gamma}^\mu_{\gamma ^{\star} N}(\vec{k}^+) 
\ket{\sigma_1\tau_1,\sigma_2\tau_2}\nonumber\\&
-\bra{s_2t_2,s_1t_1} \widehat{\Gamma}^\mu_{\gamma ^{\star} N}(\vec{k}^-) 
\ket{\sigma_1\tau_1,\sigma_2\tau_2}\,.
\end{align}

The factorization function $F^D_{n_1l_1,n_2l_2}(\vec{P}_{12})$ in
Eq.~(\ref{eq:eeNNfactorized}) can be associated with the distorted
c.m.~momentum distribution of pairs in a relative ($nl=00$) state of
the nucleus $A$
\begin{eqnarray} \label{eq:FD}
& & F^D_{n_1l_1,n_2l_2}(\vec{P}_{12}) = 4
\sum_{m_{l_1} m_{l_2}}
\Big|
\sum_{N
\Lambda M_\Lambda} \int
d\vec{R}_{12}e^{-i\vec{P}_{12}\cdot\vec{R}_{12}}\nonumber \\ &\times&
\braket{l_1m_{l_1}l_2m_{l_2}}{\Lambda M_\Lambda}
\braket{n_1l_1n_2l_2;\Lambda}{00N\Lambda ;\Lambda}_{\text{SMB}}
\nonumber\\& \times&
\mathcal{F}
^\dagger_ { \text {
FSI}}(\vec{R}_{12},\vec{R}_{12})
\psi_{N\Lambda M_\Lambda}(\sqrt{2}\vec{R}_{12})\Big|^2\, ,
\end{eqnarray}
where the factor 4 accounts for the spin degeneracy of the HO states.

\begin{figure}
  \centering
  \includegraphics[width=\columnwidth]{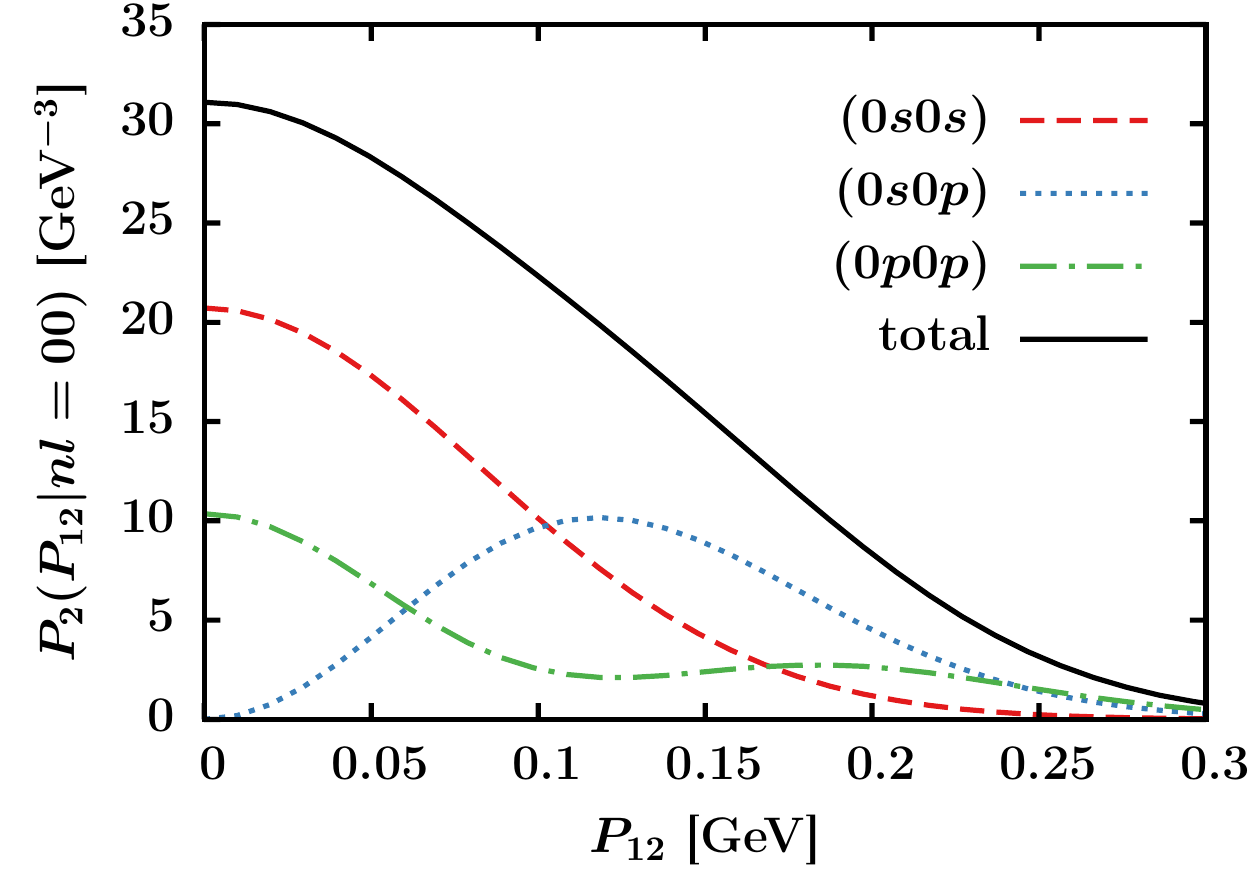}
  \caption{(Color online) The contribution of the different
    shell-model pair combinations to the $P_2\left( P_{12} | nl=00
    \right)$ for pp pairs in $^{12}$C.  }
  \label{fig:dens_com_v2}
\end{figure}

In the limit of vanishing FSIs ($\mathcal{F}_{\text{
    FSI}}\equiv 1$), one has 
\begin{multline}
  P_2(P_{12}|nl=00)  = \frac{1}{A(A-1)} \frac{3}{(2\pi)^3} \\ \times
  \sum_{n_\alpha l_\alpha n_\beta l_\beta} 
  \int d\Omega_{P_{12}}
  F^D_{n_\alpha l_\alpha, n_\beta l_\beta}(\vec{P}_{12})\; .
\end{multline}
This establishes a connection between the $A(e,e'NN)$ factorization function
and the contribution of pairs with quantum numbers $(n_1l_1n_2l_2)$ to
$P_{2} (P_{12}|nl=00)$, illustrated for pp pairs in $^{12}$C in 
Fig.~\ref{fig:dens_com_v2}.

In the naive IPM, each two-hole (2h) state
$(n_1l_1)^{-1}(n_2l_2)^{-1}$ can be associated with a sharp excitation
energy in the $A-2$ system. In reality, the 2h strength corresponding
with $(n_1l_1)^{-1}(n_2l_2)^{-1}$ extends over a wide energy range
\cite{Barbieri:2004xn}.  Current $A(e,e'pN)$ measurements are
performed at $Q^{2}$-values of the order of GeV$^2$ not allowing one
to measure cross sections for real exclusive processes as could be
done at lower $Q^2$ values
\cite{Starink2000,Ryckebusch:2003tu,Middleton:2009zd}. Accordingly,
rather than probing the individual 2h contributions to $P_2$, the
measured semi-inclusive $A(e,e'pN)$ cross sections can be linked to
the $P_2(P_{12} \mid nl=00)$ which involves a summation over the 2h
states. From Fig.~\ref{fig:dens_com_v2} it can be appreciated that in
high-resolution $A(e,e'pN)$ measurements the c.m. distribution depends
on the two-hole structure of the discrete final A-2 state
\cite{Ryckebusch:2003tu,Barbieri:2004xn}.

The $A(e,e'p)$ reaction allows one to access the
$P_{1}(\vec{k}_m,E_m)$ modulo corrections from FSIs. It is worth
stressing that there is no simple analogy for the $A(e,e'pN)$ reaction
and that a direct connection with the two-body spectral function
$P_{2}(\vec{P}_{12}, \vec{k}_{12},E_{2m})$ is by no means evident, if
not impossible.

\section{Monte Carlo simulations} \label{sec:mc}

In this section, we investigate the implications of the proposed
factorization of Eq.~(\ref{eq:eeNNfactorized}) for the $A(e,e'pp)$
opening-angle and c.m. distributions accessible in typical measurements. 
We present Monte Carlo simulations for $A(e,e'pp)$ building on the
expression (\ref{eq:eeNNfactorized}) suggesting that the magnitude of
the cross section is proportional to $P_2(P_{12}|nl=00)$. In
this section the effects of FSIs are neglected. Its impact will be the
subject of Sect.~\ref{sec:FSI}.

The data-mining effort at CLAS in Jlab
\cite{DataMining2013,Hen:2012jn} is analyzing exclusive $(e,e'pN)$ for
$^{12}$C, $^{27}$Al, $^{56}$Fe, and $^{208}$Pb for a 5.014~GeV
unpolarized electron beam \cite{DataMining2013}.  In order to
guarantee the exclusive character of the events, cuts are applied to
the leading proton: $0.62 < \frac{|\vec{p}_1|}{|\vec{q}|} < 0.96$,
$\theta_{\vec{p}_1,\vec{q}} < \unit{25}{\degree}$ and $k_{1} >
\unit{300}{\mega\electronvolt}$. To increase the sensitivity to
SRC-driven processes one imposes the kinematic constraints $x_B =
\frac{Q^{2}}{2M_N \omega} > 1.2$ and $Q^2 > 1.4$~GeV$^2$.  We have
performed $(e,e'pp)$ simulations for all 4 target nuclei.  The
electron kinematics are drawn from the measured $x_B-Q^2$
distributions.  We then generate two protons from the phase space by
adoping a reaction picture of the type (\ref{eq:eepNreaction}) whereby
we assume that one nucleon absorbs the virtual photon.  This results
in a fast leading proton $p_1(E_1,\vec{p}_1=\vec{k}_1+\vec{q})$ and a
recoil proton $p_2(E_2, \vec{p}_2 = \vec{k}_2)$, where $\vec{k}_1$ and
$\vec{k}_2$ are the initial proton momenta.  The initial c.m. momentum
$\vec{P}_{12}=\vec{k}_1+\vec{k}_2$ is drawn from the computed HO pp
pair c.m. momentum distribution $P_2(P_{12}|nl=00)$ of
Table~\ref{tab:moments}. We choose $\vec{k}_1$ along the $z$-axis and
$\vec{q}$ in the $xz$ plane.  The recoil $A-2$ nucleus can have
excitation energies between $0$ and $80$~MeV. All $A(e,e^\prime pp)$
results of this section are obtained for $10^{5}$ events which comply
with the kinematic cuts.

\begin{figure}
  \centering \includegraphics[width=\columnwidth]{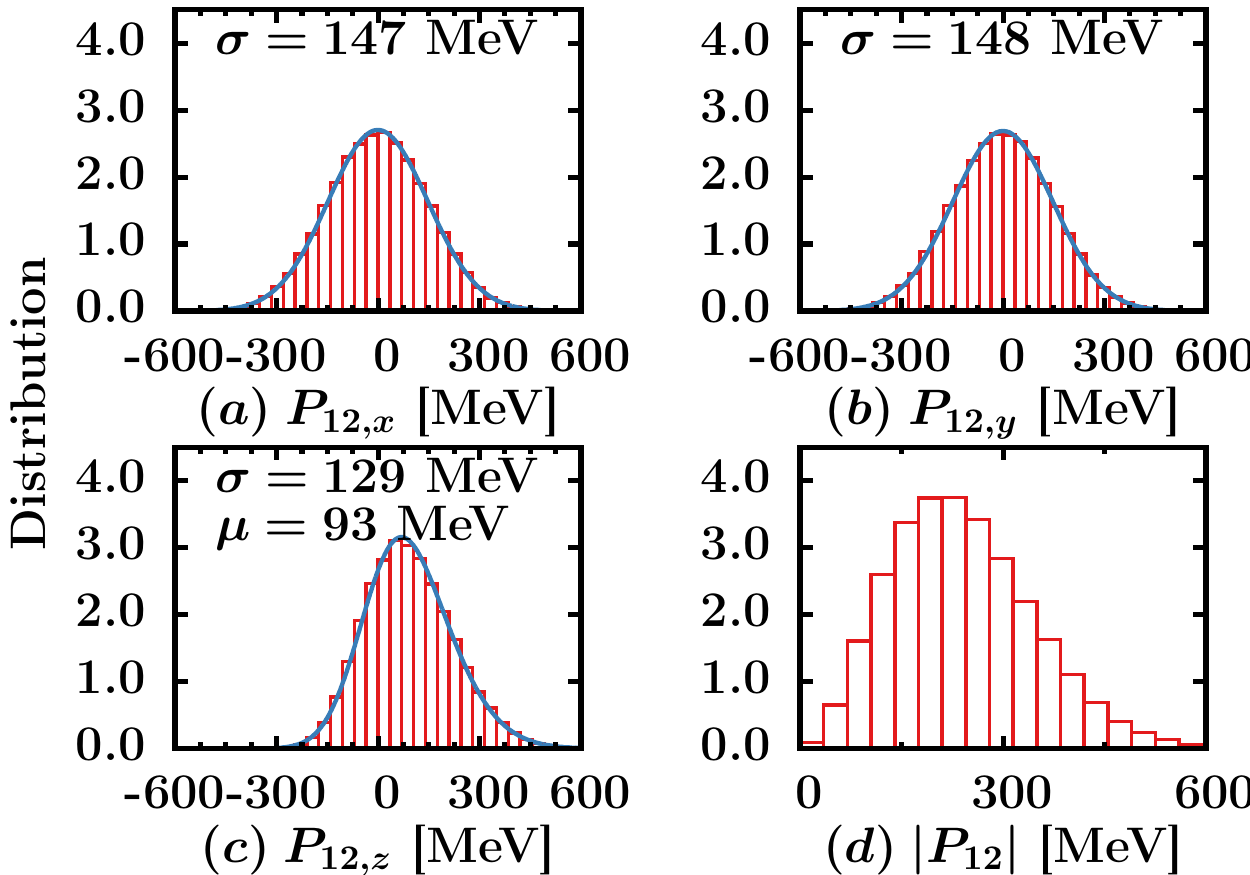}
  \caption{ (Color online) Total (bottom right) and directional pp
    c.m. distributions extracted from the $^{12}$C$(e,e'pp)$
    simulations in the CLAS kinematics described in the text. The blue
    solid line is a fit with a skew normal distribution.}
    \label{fig:MCshiftex}
\end{figure}

First, we investigate in how far the factorization function can be
addressed after applying kinematic cuts. This can be done by comparing
the input and extracted pp c.m.~distributions.
Fig.~\ref{fig:MCshiftex} shows the extracted c.m. distribution from
the simulated $^{12}$C$(e,e'pp)$ events.  The kinematic cuts have a
narrowing effect (less than 10~\%) on the distributions along the $x$-
and $y$-axis.  In addition, one observes a shift of roughly $100$~MeV
and an increase in the non-Gaussianity of the c.m.~distribution along
the $z$-axis. Similar observations have been made for the other three
target nuclei.

We now address the issue whether the extracted c.m. distributions can
provide information about the relative quantum numbers of the
pairs. To this end, we have performed simulations starting from the
assumption that the $(e,e'pp)$ cross section factorizes with $P_2 ( P_{12}
| nl)$ for various $nl$ combinations.  The results of the simulations
are summarized in Table~\ref{tab:simwidths}. The narrowing effect
attributed to the kinematic cuts is less significant for $l>0$ pairs.
Photon absorption on $l=0$ and $l=1$ pairs leads to differences in the
extracted widths of the c.m. momentum distributions of the order of
20~MeV, which leads us to conclude that high-accuracy $A(e,e'pp)$
experiments could indeed provide information about the relative
orbital angular momentum of the correlated pairs.

\begin{figure}[tbp]
  \centering
  \includegraphics[width=\columnwidth]{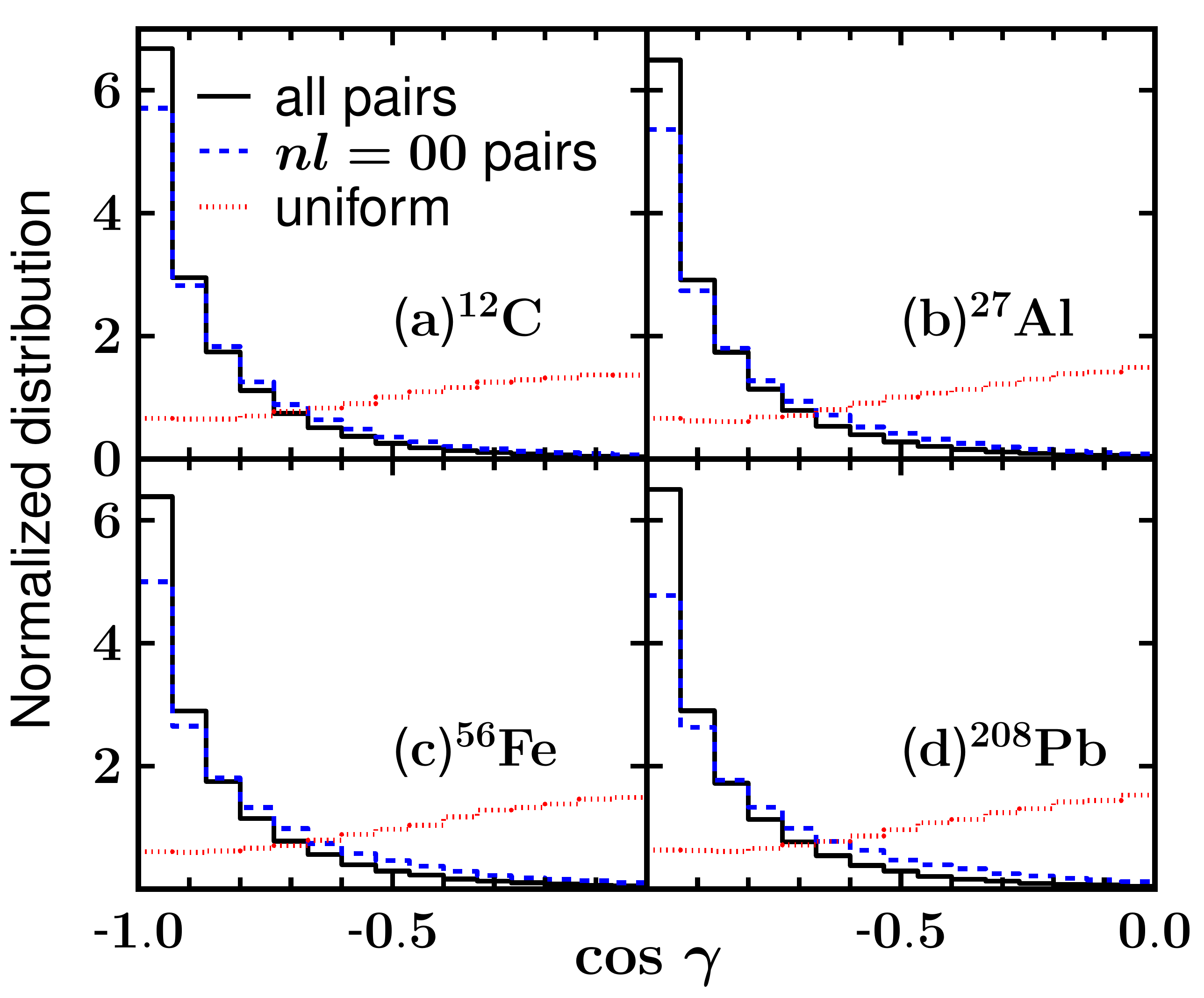}
\caption{(Color online) The opening angle distribution of the
  simulated A$(e,e'pp)$ events in the kinematics described in the
  text.  The black solid, blue dashed and red dotted line is for a
  reaction picture with an $(e,e'pp)$ cross section proportional to $ P_{2}
  (P_{12}) $, to $ P_{2} (P_{12} | nl=00) $, and to a uniform pair
  c.m. distribution.}
    \label{fig:gamma_all}
\end{figure}

Fig.~\ref{fig:gamma_all} shows the simulated opening-angle ($\gamma$)
distributions of the initial-state protons for all four target nuclei
considered.  The A$(e,e'pp)$ simulations starting from the computed
$P_2 ( P_{12} | nl=00)$ and $P_2 ( P_{12})$ provide very similar
backwardly peaked $\cos \gamma$ distributions.  The peak is not due to
the kinematic cuts as a uniform c.m. momentum distributions gives rise
to a flat $\cos \gamma$ distribution.  The shape of the simulated
$\cos \gamma$ distributions is hardly target-mass dependent. The peak
at 180 degrees in the $\cos \gamma $ distributions conforms with the
picture of correlated nucleons moving back to back with high relative and
low c.m.~momentum.

\begin{table}[tbp]
  \centering
  \begin{tabular}{cccccc}
    \hline \hline 
& $nl=00$ & $l=0$ & $l=1$ & $l=2$ & all $l$ \\
\hline 
$\sigma_{x}^i(MeV)$ & $156$ & $154$ & $135$ & $121$ & $140$ \\
$\sigma_{x}^{f}(MeV)$ & $147$ & $145$ &   $130$ &   $118$ & $134$ \\
    \hline \hline 
  \end{tabular}
  \caption{ The width of the c.m. distribution along the $x$-axis for
    pp pairs with different relative orbital momentum $l$.
    $\sigma_x^i$ is the width used as input parameter in the
    $^{12}$C$(e,e'pp)$ simulations.  The $\sigma_x^f$ is the width
    extracted after the simulation.  }
  \label{tab:simwidths}
\end{table}

\begin{figure}[tbp]
  \centering
  \includegraphics[width=\columnwidth]{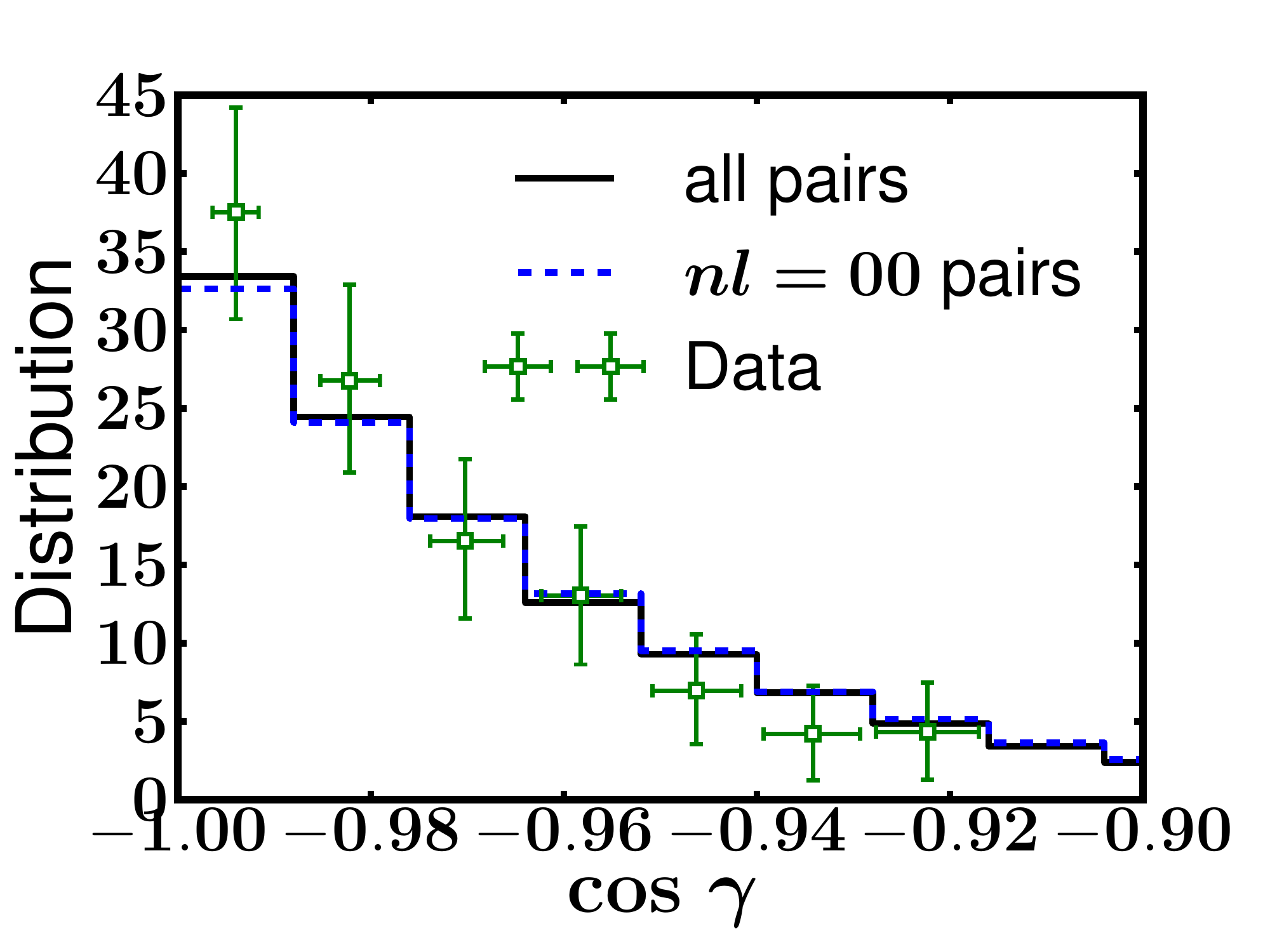}
  \caption{(Color online) The opening angle distribution of the
    $^{12}$C$(e,e'pp)$ reaction in the kinematics of
    Ref.~\cite{Shneor:2007tu}. Curve notations of
    Fig.~\ref{fig:gamma_all} are used.}
    \label{fig:gamma_HALLA}
\end{figure}

We now turn our attention to an $^{12}$C$(e,e'pp)$ measurement probing
a restricted part of phase space.  The JLab Hall-A
$^{12}$C($e,e^\prime pp$) experiment of 
Refs.~\cite{Shneor:2007tu,Subedi:2008zz}, used an incident electron beam of
$4.672$~GeV and three spectrometers.  We consider the kinematic settings with 
$\omega=0.865$~GeV, $Q^2= 2$~GeV$^2$, $x_B=1.2$
and a median missing momentum $p_m=0.55$~GeV.
Figure~\ref{fig:gamma_HALLA} shows the shapes of the simulated and
measured $\cos \gamma$ simulations.  The proposed factorization for
the $A(e,e'pp)$ cross section accounts for the shape of the measured
$\cos \gamma$ distribution. We stress that the computed pair
c.m. distributions (Table~\ref{tab:moments}) are the sole input to the
simulations.

\section{FINAL STATE INTERACTIONS} \label{sec:FSI}

In this section the impact of FSIs on the proposed factorization 
function of Eq.~(\ref{eq:eeNNfactorized}) is
investigated.  In order to keep computing times reasonable we limit
ourselves to some particular kinematic cases and introduce an
additional approximation.  We start from Eq.~(\ref{eq:FD}) for the
distorted momentum distribution $F^D_{n_1l_1,n_2l_2}(\vec{P}_{12})$
and apply the zero-range approximation
\cite{Ryckebusch:1996wc,Cosyn:2009bi} which amounts to setting
$\psi_{\alpha_1}(\vec{r}_1)\psi_{\alpha_2}(\vec{r}_2) \approx
\psi_{\alpha_1}(\vec{R}_{12})\psi_{\alpha_2}(\vec{R}_{12})$ in
Eq.~(\ref{eq:matrixHOized}). Consequently, we can write
\begin{multline} \label{eq:FD_zero}
 F^D_{n_1l_1,n_2l_2}(\vec{P}_{12}) = 4
\sum_{m_{l_1} m_{l_2}}
\Big|
 \int d\vec{R}_{12}e^{-i\vec{P}_{12}\cdot\vec{R}_{12}} \\  \times
\mathcal{F}
^\dagger_ { \text {
FSI}}(\vec{R}_{12},\vec{R}_{12})
\psi_{n_1l_1m_{l_1}}(\vec{R}_{12})\psi_{n_2l_2m_{l_2}}(\vec{R}_{12})\Big|^2\,.
\end{multline}
It is possible to derive a relativized version of this expression 
\cite{Cosyn:2009bi} 
\begin{multline}\label{eq:FD_RMSGA}
	F^D_{n_1 \kappa_1 , n_2 \kappa_2}(\vec{P}_{12}) = \\ 
	\sum_{s_1,s_2,m_1,m_2} \left| \int d\vec{R}_{12} \, e^{i  \vec{P}_{12}
\cdot \vec{R}_{12} } \bar{u}(\vec{k}_1,s_1) \psi_{n_1 \kappa_1
m_1}(\vec{R}_{12}) \right. \\ \left. \bar{u}(\vec{k}_2,s_2) \psi_{n_2 \kappa_2
m_2}(\vec{R}_{12}) \mathcal{F}_{\text{FSI}}(\vec{R}_{12},\vec{R}_{12})
\right|^{2} \, .
\end{multline}
Here, $u(\vec{k},s)$ are positive-energy Dirac spinors and
$\psi_{n\kappa m}$ are relativistic mean-field wave functions
\cite{Furnstahl:1996wv} with quantum numbers $(n, j=|\kappa|/2,
m)$. We neglect the projections on the lower components of the
plane-wave Dirac spinors.  The FSIs of the ejected pair with the
remaining $A-2$ spectators, encoded in
$\mathcal{F}_{\text{FSI}}$, can be computed in a
relativistic multiple-scattering Glauber approximation (RMSGA)
\cite{Ryckebusch:2003fc,Cosyn:2013qe}.  As the c.m. momentum is
conserved in interactions among the two ejected nucleons, we discard 
those. This approximation does not affect the shape of $F^D_{n_1
  \kappa_1 , n_2 \kappa_2}(\vec{P}_{12})$.

\begin{figure}
\includegraphics[width=\columnwidth]{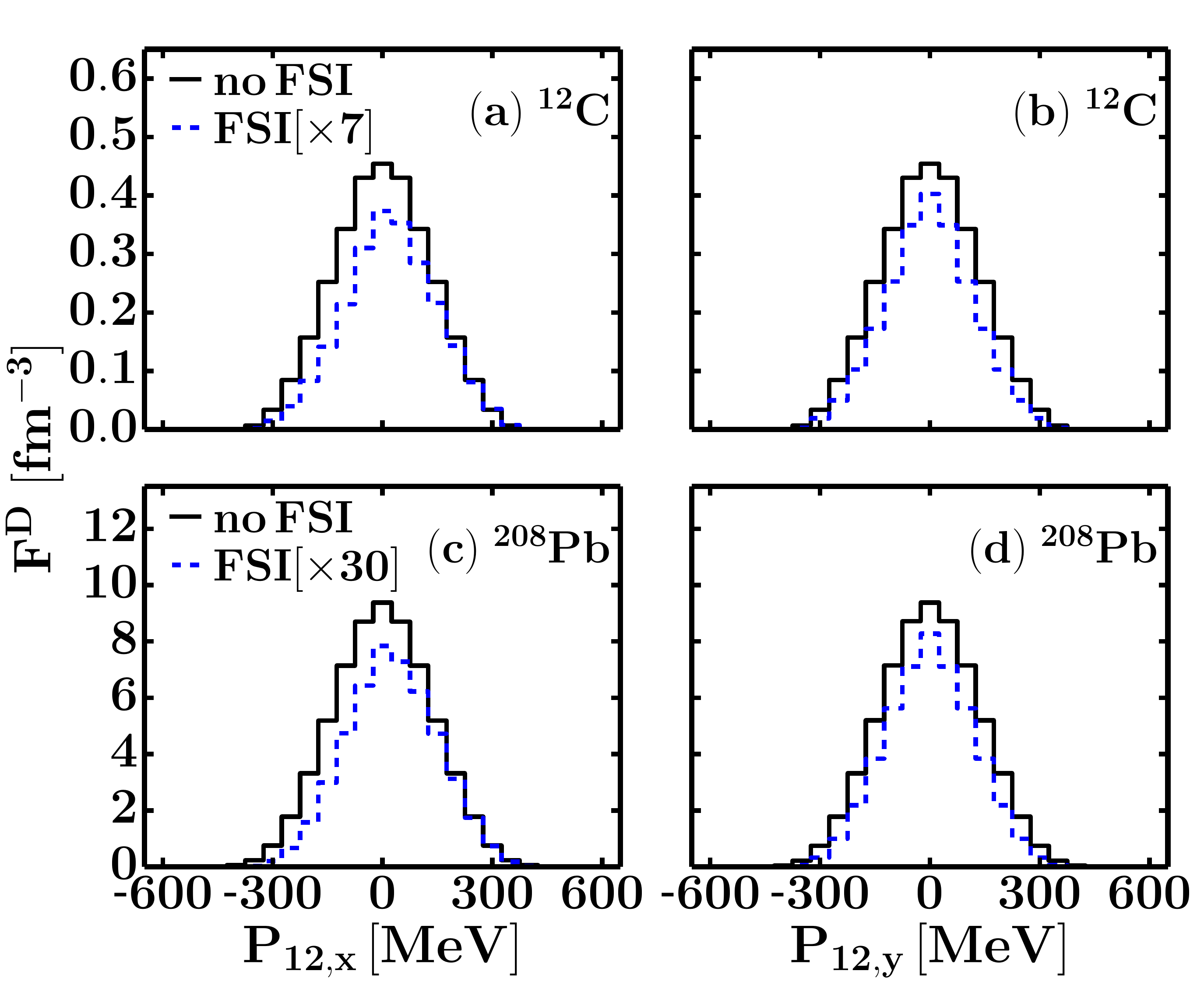}
\caption{(Color online) The two-body c.m. momentum distribution for
  $^{12}\text{C}(e,e'pp)$ (top) and $^{208}\text{Pb}(e,e'pp)$ (bottom)
  with (RMSGA) and without (no-FSI) inclusion of FSIs. We consider the
  kinematics $ |\vec{q}| = 1.4~\text{GeV}, |\vec{p}_1| = 0.82
  |\vec{q}|$ and $\theta_{\vec{p}_1,\vec{q}} = 10^{\circ}$. The FSI results
  have been multiplied by a factor of $7$ for $^{12}\text{C}(e,e'pp)$
  and by a factor of $30$ for $^{208}\text{Pb}(e,e'pp)$.}
\label{fig:FSI_proj}
\end{figure}
\begin{figure}
\includegraphics[width=\columnwidth]{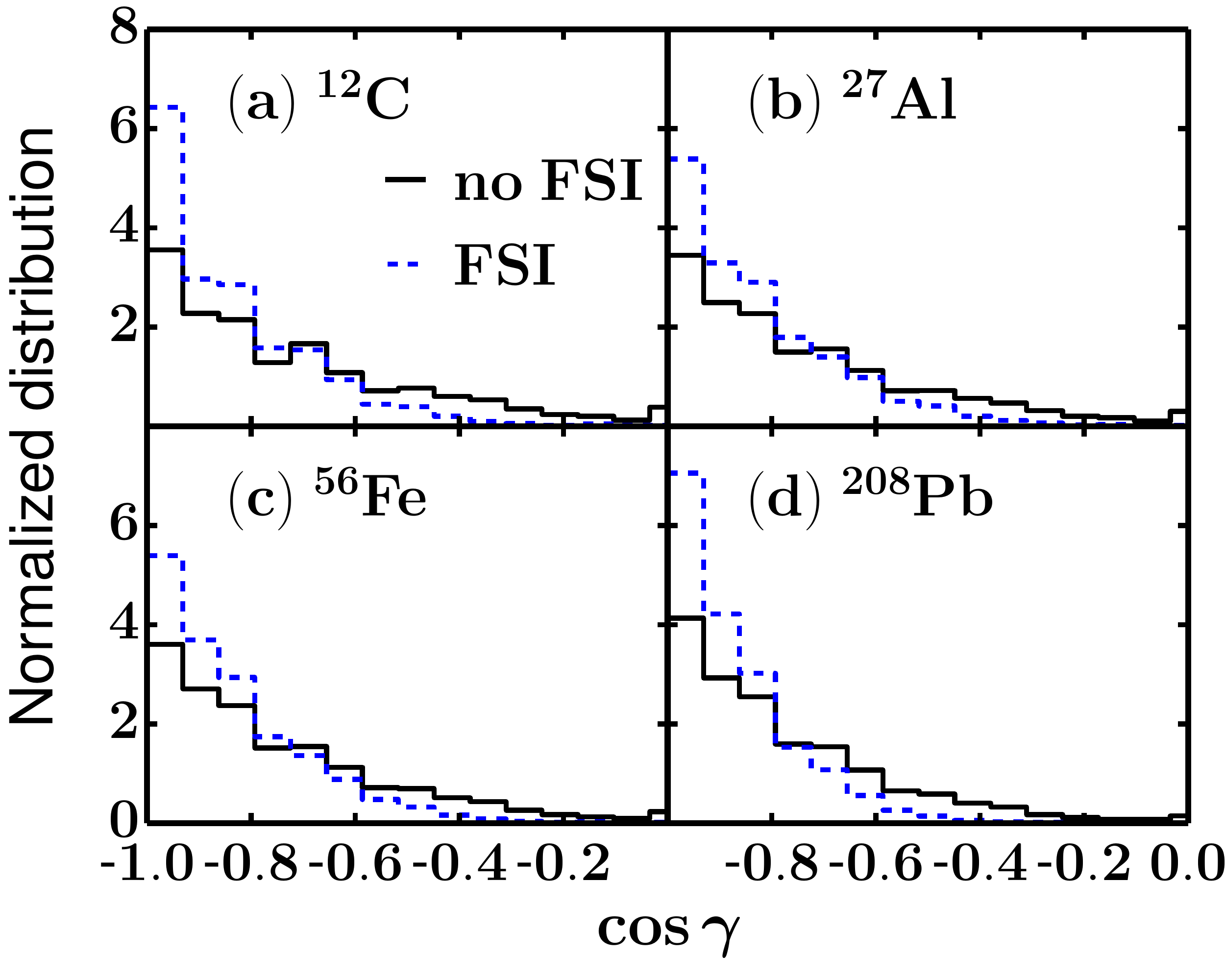}
\caption{(Color online) The normalized opening angle distributions for
  A$(e,e'pp)$ for $^{12}$C, $^{27}$Al, $^{56}$Fe and $^{208}$Pb in the
  kinematics of Fig.~\ref{fig:FSI_proj}. }
\label{fig:opening_angles_fsi}
\end{figure}

We include FSIs for the JLab data mining kinematics considered
in Sec.~\ref{sec:mc}. We have computed the distorted c.m. momentum
distribution of Eq.~(\ref{eq:FD_RMSGA}) for the kinematics that yields
the most events in the simulations of Sec.~\ref{sec:mc}: $ |\vec{q}| =
1.4~\text{GeV}, |\vec{p}_1| = 0.82 |\vec{q}|,
\theta_{\vec{p}_1,\vec{q}} = 10^{\circ}$. As in Sec.~\ref{sec:mc},
$\vec{k}_1$ lies along the $z$-axis and the $\vec{q}$ is located in
the $xz$ plane.  The results of the FSI calculations are summarized in
Figs.~\ref{fig:FSI_proj} and \ref{fig:opening_angles_fsi}.

In Fig.~\ref{fig:FSI_proj} we compare the RMSGA c.m.~momentum
distributions $F^D(\vec{P}_{12,x}) = \sum_{n_1 \kappa_1, n_2 \kappa_2}
F^D_{n_1 \kappa_1, n_2 \kappa_2}(\vec{P}_{12,x})$ and
$F^D(\vec{P}_{12,y})$ with their respective plane-wave (no-FSI)
limit. First, the FSIs are responsible for a substantial reduction of
the cross sections: a factor of about 7 for carbon and about 30 in
lead. The effects of FSIs on the shape of $F^D(\vec{P}_{12})$,
however, are rather modest.  Gaussian fits to the
$F^D(\vec{P}_{12,i=x,y})$ result in widths which are less than 10\%
smaller than in the plane-wave limit. The effects of FSIs on the shape
of the c.m. distributions in Fig.~\ref{fig:FSI_proj} can be
qualitatively understood considering that the nucleons undergoing FSIs
are slowed down on average: $( \vec{p}_1,\vec{p}_2 )
\xrightarrow{\text{FSI}} \zeta \left( \vec{p}_1, \vec{p}_2 \right)$
with $0 < \zeta \le 1 $. It is straightforward to show that for the
adopted conventions this results in $P_{12,x} \rightarrow \zeta
P_{12,x} - (1- \zeta )p_{1,x} $, and $P_{12,y} \rightarrow \zeta
P_{12,y}$. This explains the observed contraction and shift to the
right in the $P_{12,x}$ distribution, and the contraction of the
$P_{12,y}$ distributions.

The effect of FSIs on the shape of the normalized opening angle
distributions is studied in Fig.~\ref{fig:opening_angles_fsi} for four
target nuclei.  It is clear that they become even more forwardly
peaked after including FSIs.
 

\section{Summary} \label{sec:concl}

Summarizing, we have shown that in the plane-wave limit the
factorization function for the exclusive SRC-driven $A(e,e'pN)$
reaction is the conditional c.m.~distribution $P_2(P_{12}|nl=00)$ for
pN pairs in a nodeless relative state with a vanishing orbital
momentum. We have illustrated that in a two-body cluster expansion the
correlated part of the momentum distribution originates mainly from
correlation operators acting on IPM pairs with $(nl=00)$ quantum
numbers, supporting the assumptions underlying the proposed
factorization of the $A(e,e'pN)$ reaction.  Numerical
calculations indicate that the $P_2(P_{12}|nl=00)$ has a wider
distribution than the unconditional $P_2(P_{12})$ one. An important
implication of the proposed factorization is that the mass dependence
of the $A(e,e'pp)$ and $A(e,e'pn)$ cross section is predicted to be
much softer than $\frac {Z(Z-1)}{2}$ and $NZ$ respectively.

We have examined the robustness of the proposed factorization of
the two-nucleon knockout cross sections against kinematic cuts and
FSIs. Both mechanisms modestly affect the shape of the
c.m.~distributions which leads us to conclude that they can be
accessed in $A(e,e'pN)$ measurements. The FSIs bring about a
mass-dependent reduction of the cross sections which is of the order
of 10 for carbon and 30 for lead.

\subsection*{ACKNOWLEDGMENTS}
The authors wish to thank Or Hen, Eli Piasetzky, and Larry Weinstein
for stimulating discussions and suggestions.  This work is supported
by the Research Foundation Flanders (FWO-Flanders) and by the
Interuniversity Attraction Poles Programme P7/12 initiated by the
Belgian Science Policy Office. The computational resources (Stevin
Supercomputer Infrastructure) and services used in this work were
provided by Ghent University, the Hercules Foundation and the Flemish
Government.

\bibliography{CMpaper}
\end{document}